\documentclass[twocolumn,showpacs,amsmath,amssymb,superscriptaddress,footinbib]{revtex4}

\usepackage{amsfonts}
\usepackage{amsmath}
\usepackage{aas_macros}
\usepackage{graphicx}
\usepackage{natbib}
\usepackage{color}

\def\lsim{~\rlap{$<$}{\lower 1.0ex\hbox{$\sim$}}}
\def\bsim{~\rlap{$>$}{\lower 1.0ex\hbox{$\sim$}}}

\def\hmpc{\ {\rm {\it h}^{-1}Mpc}}

\def\hmsun{\ {\rm M_\odot/{\it h}}}

\def\la{\langle}
\def\ra{\rangle}

\def\ln{{\rm ln}}
\def\tr{{\rm tr}}
\def\det{{\rm det}}

\def\mathbi#1{\textbf{\em #1}}

\def\ov{\overline}

\def\dsc{\delta_{\rm sc}}

\def\veta{{\boldsymbol\eta}}
\def\vzeta{{\boldsymbol\zeta}}

\def\czeta{\smash[b]{\ov{\zeta}}}
\def\evh{\mathrm{\hat{\bf{e}}}}
\def\kvh{\mathrm{\hat{\bf{k}}}}

\def\rvh{\mathrm{\hat{\bf{r}}}}

\def\kh{\mathit{\hat{k}}}

\def\rh{\mathit{\hat{r}}}
\def\ve{\mathrm{\bf e}}
\def\vk{\mathrm{\bf k}}

\def\vr{\mathrm{\bf r}}

\def\vu{\mathrm{\bf u}}
\def\vv{\mathrm{\bf v}}

\def\vx{\mathrm{\bf x}}
\def\vy{\mathrm{\bf y}}

\def\vbb{\mathrm{B}}
\def\vcc{\mathrm{C}}

\def\vmm{\mathrm{M}}

\def\grad{\mathbi{$\nabla$}}

\def\bias#1{\mathfrak{\tilde{b}}_{\rm{#1}}}
\def\npk{n_{\rm pk}}
\def\bnpk{\bar{n}_{\rm pk}}
\def\dpk{\delta_{\rm pk}}

\def\xpk{\xi_{\rm pk}}

\def\wi#1{\widehat{#1}}
\def\nesp{\bar{n}_{\rm ESP}}

\def\apjl{Astrophys.\ J.\ Lett.}

\def\aap{Astron.\ Astrophys.}

\def\apjs{Astrophys.\ J. Supp.}

\newcommand{\beq}{\begin{equation}}
\newcommand{\eeq}{\end{equation}}
\newcommand{\beqa}{\begin{eqnarray}}
\newcommand{\eeqa}{\end{eqnarray}}

\begin{document}

\title[A local bias approach to the clustering of discrete density peaks]
      {A local bias approach to the clustering of discrete density peaks}

\author{Vincent Desjacques} \email{Vincent.Desjacques@unige.ch}
\affiliation{D\'epartement de Physique Th\'eorique and Center for Astroparticle Physics (CAP)
Universit\'e de Gen\`eve, 24 quai Ernest Ansermet, CH-1211 Gen\`eve, Switzerland}

%%%%%%%%%%%%%%%%%%%%%%%%%%%%%%%%%%%%%%%%%%%%%%%%%%%%%%%

\begin{abstract}

Maxima of the linear density field form a point process that can be used to understand the 
spatial distribution of virialized halos that collapsed from initially overdense 
regions. 
However, owing to the peak constraint, clustering statistics of discrete density peaks are 
difficult to evaluate. 
For this reason, local bias schemes have received considerably more attention in the
literature thus far. In this paper, we show that the 2-point correlation function of maxima of
a homogeneous and isotropic Gaussian random field can be thought of, up to second order at least,
as arising from a local bias expansion formulated in terms of rotationally invariant variables.
This expansion relies on a unique smoothing scale, which is the Lagrangian radius of dark matter 
halos. 
The great advantage of this local bias approach is that it circumvents the difficult computation 
of joint probability distributions.
We demonstrate that the bias factors associated with these rotational invariants can be computed 
using a peak-background split argument, in which the background perturbation shifts the 
corresponding probability distribution functions.
Consequently, the bias factors are orthogonal polynomials averaged over those spatial locations 
that satisfy the peak constraint. 
In particular, asphericity in the peak profile contributes to the clustering at quadratic and 
higher order, with bias factors given by generalized Laguerre polynomials. 
We speculate that our approach remains valid at all orders, and that it can be extended to 
describe clustering statistics of any point process of a Gaussian random field. Our results will
be very useful to model the clustering of discrete tracers with more realistic collapse 
prescriptions involving the tidal shear for instance.

\end{abstract}

\pacs{98.80.-k,~98.65.-r,~95.35.+d,~98.80.Es}
\maketitle

\setcounter{footnote}{0}

\section{Introduction}
\label{sec:intro}

In the biasing scenario introduced by \cite{1984ApJ...284L...9K}, virialized halos form 
out of initially overdense regions with a linear density (extrapolated to the redshift of 
interest) equal to $\delta_c\approx 1.686$. Since then, this picture has received 
considerable support from observational data. Even though dark matter halos are extended
objects, they form a spatial point process as far as their clustering is concerned. 
However, this essential feature has remained elusive in most theoretical descriptions of 
halo clustering, which assume that halos are a Poisson sampling of a more fundamental, 
continuous halo density field $\delta_{\rm h}(\vx)$. 

The peak formalism first proposed by \cite{1985MNRAS.217..805P,1986ApJ...304...15B} in a 
cosmological context is interesting because it is a well-behaved point process. In this 
approach, virialized halos are associated with maxima of the initial density field. The
displacement from their initial (Lagrangian) to final (Eulerian) position can be computed
upon assuming phase space conservation \cite{2010PhRvD..82j3529D}. Clustering statistics
of these discrete density peaks display many of the features present in measurements of 
halo clustering extracted from N-body simulations. 
In particular, discrete density peaks exhibit a $k$-dependent linear bias factor
\cite{1999ApJ...525..543M,2008PhRvD..78j3503D}, small-scale exclusion 
\cite{1989MNRAS.238..293L,1989MNRAS.238..319C}, and a linear velocity bias 
\cite{2010PhRvD..81b3526D} etc. Some of these predictions have recently been tested in 
numerical simulations \cite{2011MNRAS.413.1961L,2012MNRAS.421.3472E}: peaks of the linear
density field appear to provide a good approximation to the formation sites of dark matter
halos with $M\gtrsim M_\star$. 

However, despite recent progress towards the computation of peak clustering statistics
\cite{2010PhRvD..82j3529D} and a formulation of peak theory within the excursion set 
formalism \cite{2012MNRAS.426.2789P,2012arXiv1210.1483P},
discrete density peaks lack a clear connection with the more conventional local bias 
schemes \cite{1993ApJ...413..447F}, in which halos are approximated as a continuous field. 
Furthermore, while in the local bias model the computation of halo correlation functions is 
straightforward (though there are ambiguities regarding the filtering scale etc.), in the 
peak formalism calculations are particularly tedious owing to the peak constraint 
\cite{1986ApJ...304...15B,1995MNRAS.272..447R,2008PhRvD..78j3503D,2010PhRvD..82j3529D}.
In the most comprehensive analysis thus far, ref.~\cite{2010PhRvD..82j3529D} succeeded in
computing the peak 2-point correlation $\xpk(r)$ up to second order, including the Zel'dovich
displacement. They showed that some of the first- and second-order contributions could be 
obtained from a peak-background split formulated in terms of conditional mass functions.
In contrast to most analytic models of halo clustering, which assume that the 
($k$-independent) bias coefficients are the peak-background split biases, they derived this
equivalence from first principles. However, they could not determine the physical origin
of the other second-order contributions.
Moreover, the peak constraint is clearly too simplistic to describe the clustering of low
mass halos. In this mass range, one should consider more elaborated constraints involving 
the tidal shear etc. In this regards, it would be very desirable to find a simpler way of 
computing the correlation functions of generic point processes of a (Gaussian) random field.

In this paper, we suggest a simple, physically motivated prescription based on the 
peak-background split to compute the correlation functions of generic point processes driven
by homogeneous and isotropic Gaussian random fields. We argue that clustering statistics of
such point processes can be reduced to the evaluation of correlators of an {\it effective} 
continuous overdensity which, in the case of discrete peaks, is a function of the local 
(smoothed) mass density field and its derivatives. Our approach combines in a single coherent 
picture peak theory, peak-background split, local bias and the excursion set framework. 
For sake of clarity, we will focus on the 2-point correlation function of initial density 
peaks as computed in \cite{2010PhRvD..82j3529D} to explain the fundamentals of our approach.
 
The paper is organized as follows. Sec.~\ref{sec:pkcorr} furnishes a brief summary of 
clustering in peak theory. Sec.~\ref{sec:physical} is the central Section of the paper, where 
we present the connection between rotational invariants, peak-background split and a local 
peak bias prescription. Finally, Sec.\ref{sec:discussion} discusses the implications of our 
findings.

\section{Correlation functions for density peaks}
\label{sec:pkcorr}

We begin with a short recapitulation of the computation of correlation functions in the peak 
formalism.
Let $\delta_s$ be the linear mass density field smoothed on scale $R_s$ with a spherically
symmetric filter. For convenience, we work with the normalized variables 
$\nu(\vx)\equiv \frac{1}{\sigma_0}\delta_s(\vx)$,
$\eta_i(\vx)\equiv \frac{1}{\sigma_1}\partial_i\delta_s(\vx)$ and
$\zeta_{ij}(\vx)\equiv \frac{1}{\sigma_2}\partial_i\partial_j\delta_s(\vx)$.
Here, $\nu$ is the peak height or significance, and 
\begin{equation}
\sigma_n^2(R_s) \equiv \frac{1}{2\pi^2}\int_0^\infty\!\!dk\,k^{2(n+1)}\, P_s(k)\;.
\label{eq:mspec}
\end{equation} 
are moments of the power spectrum $P_s(k)=\la|\delta_s(\vk)|^2\ra$. A Gaussian filter is 
frequently adopted to ensure convergence of all the spectral moments $\sigma_n$, but one should 
bear in mind that the peak height associated with dark matter halos is always computed with 
a tophat filter (see \cite{2012arXiv1210.1483P} for details). The first few spectral moments
$\sigma_n$ can be combined into a dimensionless spectral width 
$\gamma_1=\sigma_1^2/(\sigma_0\sigma_1$ that takes values between zero and unity. $\gamma_1$
reflects the range over which the smoothed power spectrum $P_s(k)$ is significant, i.e.
$\gamma_1\approx 1$ for a sharply peaked power spectrum whereas $\gamma_1\approx 0$ for a 
power spectrum that covers a wide range of wavenumbers.

Correlations of density maxima of $\delta_s$ can be evaluated using the Kac-Rice formula 
\citep{Kac1943,Rice1945}. The trick is to Taylor-expand $\eta_i(\vx)$ around the position 
$\vx_{\rm pk}$ of a local maximum. As a result, the number density of (BBKS) peaks of height 
$\nu'$ at position $\vx$ in the smoothed density field $\delta_s$ can be expressed in terms 
of the field $\delta_s$ and its derivatives:
\begin{align}
\label{eq:npkx}
\npk(\nu',R_s,\vx) &\equiv
\frac{3^{3/2}}{R_\star^3}|\det\zeta(\vx)|\,
\delta_D\!\left[\veta(\vx)\right]\,
\theta_H\!\left[\lambda_3(\vx)\right] \\
&\quad \times \delta_D\!\left[\nu(\vx)-\nu'\right] \nonumber \;,
\end{align}
where $R_\star \equiv \sqrt{3} (\sigma_1/\sigma_2)$ is the characteristic radius of a peak 
(and {\it not} the interpeak distance). 
The three-dimensional Dirac distribution $\delta_D\!(\veta)$ ensures that all extrema are 
included. 
The factors of theta function $\theta_H(\lambda_3)$, where $\lambda_3$ is the lowest 
eigenvalue of the shear tensor $\zeta_{ij}$, and the Dirac delta $\delta_D\!(\nu-\nu')$ 
further restricts the set to density maxima of the desired significance $\nu'$.  

The (disconnected) $N$-point correlations $\rho_{\rm pk}^{(N)}$ (or joint intensities) of 
density maxima are defined as the ensemble averages of products of $\npk(\nu,R_s;\vx)$,
\begin{multline}
\label{eq:Nptpk}
\rho_{\rm pk}^{(N)}\!(\nu,R_s,\vx_1,\dots,\vx_N) \\ 
\equiv\Bigl\langle\npk(\nu,R_s,\vx_1)\times\dots\times\npk(\nu,R_s,\vx_N)\Bigr\rangle \;.
\end{multline}
For the Gaussian initial conditions considered here, multivariate normal distribution are 
assumed to perform the ensemble average. 
In the particular case $N=1$, $\la\npk(\nu,R_s,\vx)\ra=\bnpk(\nu,R_s)$ is the average, 
differential number density of peaks of height $\nu$ identified on the filtering scale $R_s$ 
\citep{1986ApJ...304...15B},
\begin{align}
\bnpk(\nu,R_s) &= \frac{1}{(2\pi)^2 R_\star^3}\,e^{-\nu^2/2}\,
G_0^{(1)}\!(\gamma_1,\gamma_1\nu) \\ 
&= \frac{e^{-\nu^2/2}}{\sqrt{2\pi}} \left(\frac{1}{V_\star}\right)
G_0^{(1)}\!(\gamma_1,\gamma_1\nu) \nonumber \;.
\label{eq:bnpk}
\end{align}
In the last equality, $V_\star=(2\pi)^{3/2}R_\star^3$ is the typical 3-dimensional extent of 
a density peak \cite{2012MNRAS.426.2789P}.
The functions $G_n^{(\alpha)}(\gamma_1,\gamma_1\nu)$ are defined in Appendix \ref{sec:uvw}. In 
particular, the ratio $G_k^{(1)}/G_0^{(1)}$ is equal to the $k$th moment $\ov{u^k}$ of the peak 
curvature $u$.
Similarly, the reduced 2-point correlation function for maxima of a given significance separated 
by a  distance $r=|\vr|=|\vx_2-\vx_1|$ is
\begin{equation}
\xpk(\nu,R_s,r)=\frac{\rho_{\rm pk}^{(2)}\!(\nu,R_s,r)}{\bnpk^2(\nu,R_s)}-1\;,
\label{eq:2ptkacrice}
\end{equation}
Notice that, in $\rho_{\rm pk}^{(2)}$, we have ignored the shot-noise term $\bnpk\delta_D(\vx_2-\vx_1)$ 
that arises from the self-pairs as it matters only at zero-lag (in the peak power spectrum 
however, this contributes a constant Poisson noise $1/\bnpk$ at all wavenumbers).

The calculation of Eq.(\ref{eq:2ptkacrice}) at second order in the mass correlation and its 
derivatives is quite tedious \cite{1986ApJ...304...15B,1995MNRAS.272..447R,2010PhRvD..82j3529D}
because one must evaluate the joint probability distribution for the 10-dimensional vector of 
variables $\vy_\alpha^\top=(\eta_i(\vx_\alpha),\nu(\vx_\alpha),\zeta_A(\vx_\alpha))$ at two 
different spatial locations $\vx_\alpha=\vx_1$ and $\vx_2$,
i.e. a total of 20 variables. Here, the components $\zeta_A, A=1,\cdots,6$ symbolize the 
independent entries $ij=11,22,33,12,13,23$ of $\zeta_{ij}$.
Fortunately, as was shown in \cite{2010PhRvD..82j3529D}, most of the terms nicely combine 
together, so that the final result can be recast into the compact expression
\begin{widetext}
\begin{align}
\label{eq:xpkeasy}
\xpk(\nu,R_s,r) &= \bigl(\bias{I}^2\xi_0^{(0)}\bigr) +\frac{1}{2}
\bigl(\xi_0^{(0)}\bias{II}^2\xi_0^{(0)}\bigr)-\frac{3}{\sigma_1^2}
\bigl(\xi_1^{(1/2)}\bias{II}\xi_1^{(1/2)}\bigr)
-\frac{5}{\sigma_2^2}\bigl(\xi_2^{(1)}\bias{II}\xi_2^{(1)}\bigr)
\biggl(1+\frac{2}{5}\partial_\alpha\ln G_0^{(\alpha)}\!(\gamma_1,\gamma_1\nu)
\Bigl\rvert_{\alpha=1}\biggr) \nonumber \\
&\quad + \frac{5}{2\sigma_2^4}\Bigl[\bigl(\xi_0^{(2)}\bigr)^2+\frac{10}{7}
\bigl(\xi_2^{(2)}\bigr)^2+\frac{18}{7}\bigl(\xi_4^{(2)}\bigr)^2\Bigr]
\biggl(1+\frac{2}{5}\partial_\alpha\ln G_0^{(\alpha)}\!(\gamma_1,\gamma_1\nu)
\Bigl\rvert_{\alpha=1}\biggr)^2 \nonumber \\ 
&\quad +\frac{3}{2\sigma_1^4}\Bigl[\bigl(\xi_0^{(1)}\bigr)^2
+2\bigl(\xi_2^{(1)}\bigr)^2\Bigr]+\frac{3}{\sigma_1^2\sigma_2^2}
\Bigl[3\bigl(\xi_3^{(3/2)}\bigr)^2+2\bigl(\xi_1^{(3/2)}\bigr)^2\Bigr]\;.
\end{align}
\end{widetext}
The functions $\xi_\ell^{(n)}\!(r)$ are quantities analogous to $\sigma_n^2$ but defined for a 
finite separation $r$,
\begin{equation}
\xi_\ell^{(n)}\!(R_s,r)= \frac{1}{2\pi^2}\int_0^\infty\!\! dk\,
k^{2(n+1)} P_s(k)\; j_\ell(kr)\;,
\label{xielln}
\end{equation}
where $j_\ell(x)$ are spherical Bessel functions.
In the right-hand side of Eq.(\ref{eq:xpkeasy}), all the correlations depend on the filtering 
scale and the separation. However, the first line contains terms involving the first and second 
order peak bias parameters $\bias{I}$ and $\bias{II}$ (to be defined shortly), the second line 
retains a $\nu$-dependence through the function 
$1+(2/5)\partial_\alpha\ln G_0^{(\alpha)}\!(\gamma_1,\gamma_1\nu)|_{\alpha=1}$ solely, whereas 
the last two terms in the right-hand side depend on the separation $r$ (and $R_s$) only. 
Hence, unlike standard local bias expansions (Eulerian or Lagrangian), the peak 2-point 
correlation also exhibits quadratic terms linear in the second-order bias $\bias{II}$.
These terms involve derivatives of the linear mass correlation $\xi_0^{(0)}$ and, therefore,
vanish at zero lag. Clearly, they arise because the peak correlation also depends on the 
statistical properties of $\eta_i$ and $\zeta_{ij}$.

Ref.~\cite{2010PhRvD..82j3529D} also showed that the Lagrangian peak bias factors 
$\bias{N}(k_1,\dots,k_N)$ can be constructed upon averaging over the peak curvature products 
of $b_\nu$ and $b_u$, where
\begin{align}
\label{eq:bv}
b_\nu(\nu,R_s) &=
\frac{1}{\sigma_0}\left(\frac{\nu-\gamma_1 u}{1-\gamma_1^2}\right)\;, \\
b_u(\nu,R_s) &=
\frac{1}{\sigma_2}\left(\frac{u-\gamma_1\nu}{1-\gamma_1^2}\right)\;.
\label{eq:bu}
\end{align}
For peak of significance $\nu$ on the smooting scale $R_s$, the first order bias $\bias{I}$ 
is defined as the Fourier space multiplication (we omit the dependence on $R_s$ and $\nu$ for 
shorthand convenience) \cite{2008PhRvD..78j3503D}
\begin{gather}
\bias{I}(k)=b_{10} + b_{01} k^2 \\
\mbox{where} \qquad  b_{10}=\bar{b}_\nu \qquad \mbox{and} \qquad b_{01}=\bar{b}_u 
\nonumber\;.
\label{eq:bnII}
\end{gather}
The overline designates the average over the peak curvature. $b_{01}$ can be quite 
large for moderate peak heights. In the high peak limit $\nu\gg 1$ however, it is negligible 
so that $\bias{I}(k)$ is nearly scale-independent (like in local bias models).
Similarly, the Fourier space expression of the second order peak bias $\bias{II}$ is 
\cite{2010PhRvD..82j3529D}
\begin{equation}
\label{eq:bnII}
\bias{II}(k_1,k_2)= b_{20}+b_{11}\bigl(k_1^2+k_2^2\bigr)+b_{02}\,k_1^2 k_2^2\;,
\end{equation}
where $k_1$ and $k_2$ are wavemodes and the $k$-independent coefficients $b_{20}$, $b_{11}$ 
and $b_{02}$ are
\begin{align}
b_{20}(\nu,R_s) &\equiv
\ov{b_{\nu}^2}-\frac{1}{\sigma_0^2\left(1-\gamma_1^2\right)} \label{eq:bvv} \\ 
&=
\frac{1}{\sigma_0^2}\left[
\frac{\nu^2-2\gamma_1\nu\bar{u}+\gamma_1^2\ov{u^2}}
{\left(1-\gamma_1^2\right)^2}
-\frac{1}{\left(1-\gamma_1^2\right)}\right] \nonumber \\
b_{11}(\nu,R_s) &\equiv
\ov{b_{\nu}b_u}+\frac{\gamma_1^2}{\sigma_1^2\left(1-\gamma_1^2\right)} \label{eq:bvu} \\
&= \frac{1}{\sigma_0\sigma_2}\Biggl[\frac{\left(1+\gamma_1^2\right)
\nu\bar{u}-\gamma_1\bigl[\nu^2+\ov{u^2}\bigr]}
{\left(1-\gamma_1^2\right)^2} \nonumber \\ 
& \qquad +\frac{\gamma_1} {\left(1-\gamma_1^2\right)}\Biggr] \nonumber \;,
\end{align}
and
\begin{align}
b_{02}(\nu,R_s) &\equiv
\ov{b_u^2}-\frac{1}{\sigma_2^2\left(1-\gamma_1^2\right)} \label{eq:buu} \\
& = \frac{1}{\sigma_2^2}\left[\frac{\ov{u^2}-2\gamma_1\nu\bar{u}
+\gamma_1^2\nu^2}{\left(1-\gamma_1^2\right)^2}
-\frac{1}{\left(1-\gamma_1^2\right)}\right] \nonumber \;.
\end{align}
By definition, $\bias{II}^m$ acts on the functions $\xi_{\ell_1}^{(n_1)}(r)$ and 
$\xi_{\ell_2}^{(n_2)}(r)$ as follows:
\begin{multline}
\bigl(\xi_{\ell_1}^{(n_1)}\bias{II}^m\xi_{\ell_2}^{(n_2)}\bigr)
\equiv\frac{1}{4\pi^4}\int_0^\infty\!\!dk_1\int_0^\infty\!\!dk_2\,
k_1^{2(n_1+1)}k_2^{2(n_2+1)} \\ \times \bias{II}^m(k_1,k_2)P_s(k_1) P_s(k_2)
j_{\ell_1}(k_1 r) j_{\ell_2}(k_2 r)\;.
\end{multline}
As pointed out by \cite{2010PhRvD..82j3529D}, the piece 
$\bias{I}^2\xi_0^{(0)}+(1/2)\xi_0^{(0)}\bias{II}^2\xi_0^{(0)}$ can be thought of as arising 
from the continuous, deterministic, local bias relation
\begin{align}
\label{eq:olddpk}
\dpk(\vx) &= b_{10}\delta_s(\vx) - b_{01} \nabla^2\delta_s(\vx) 
+ \frac{1}{2} b_{20}\delta_s^2(\vx) \\ 
& \quad  - b_{11}\delta_s(\vx)\nabla^2\delta_s(\vx) 
+ \frac{1}{2} b_{02}\bigl[\nabla^2\delta_s(\vx)\bigr]^2 \nonumber \;,
\end{align} 
where the bias factors $b_{ij}$ are peak-background split bias factors that follow from 
expanding the conditional peak number density in a series in the small background density 
perturbation $\delta_l$. 
This expansion is local in the sense that, except for the filtering, it involves quantities 
evaluated at $\vx$ solely. However, an essential difference with the widespread local bias 
model \cite{1993ApJ...413..447F} is the fact that, when computing the ensemble average 
$\la\dpk(\vx_1)\dpk(\vx_2)\ra$, we must ignore all powers of zero-lag moments (such as, 
e.g., $\sigma_0^4$ in $\la\delta_s^2(\vx_1)\delta_s^2(\vx_2)\ra$) to recover $\xpk(r)$ since 
the latter does not exhibit such contributions (this 'no zero-lag requirement' also arises 
in the derivation of the 'renormalized' bias parameters of \cite{2012arXiv1212.0868S}). 
All the terms in Eq.(\ref{eq:olddpk}) are of course invariant under rotations since 
$\dpk(\vx)$ transforms as a scalar under rotations. 
Clearly however, this series expansion is not the most generic Lagrangian expansion we may 
conceive of (see, e.g., \cite{2012arXiv1207.7117S} non nonlocal Lagrangian bias).

Notwithstanding these results, \cite{2010PhRvD..82j3529D} did not succeed in finding a 
physical interpretation of the other second-order terms in the right-hand side of 
Eq.(\ref{eq:xpkeasy}), even though it was pretty clear that they -- at least partially -- 
arise from coupling involving the components of the gradient $\eta_i$ and the hessian 
$\zeta_{ij}$.

\section{A intuitive interpretation of $\xpk(r)$}
\label{sec:physical}

In this Section, we propose an intuitive, physically motivated  explanation of 
Eq.~(\ref{eq:xpkeasy}) 
that is grounded in the peak-background split argument \cite{1984ApJ...284L...9K}. We begin 
with a brief introduction to the helicity basis, which was used in \cite{2010PhRvD..82j3529D} 
to compute probability distributions of the density field and its derivatives at two 
different spatial locations.

\subsection{Probability density in the helicity basis}
\label{sec:helicity}

The 2-point correlation function of initial density peaks is the ensemble average of 
$\npk(\vy_1)\npk(\vy_2)$ over the joint probability density $P_2(\vy_1,\vy_2;r)$, where 
$\vy_\alpha\equiv\vy(\vx_\alpha)$ are the values of the field and its derivatives at position 
$\vx_\alpha$. In what follows, $\npk(\vy)$ will also designate Eq.(\ref{eq:npkx}).
Following \cite{2010PhRvD..82j3529D}, we can decompose the variables 
$\vy_\alpha^\top=(\eta_i(\vx_\alpha),\nu(\vx_\alpha),\zeta_A(\vx_\alpha))$ that appear in 
the joint probability density $P_2(\vy_1,\vy_2;r)$ in the helicity basis $(\ve_+,\rvh,\ve_-)$, 
where
\begin{equation}
\ve_+\equiv\frac{i\evh_\phi-\evh_\theta}{\sqrt{2}},~~~ \rvh\equiv
\vr/r,~~~ \ve_-\equiv\frac{i\evh_\phi+\evh_\theta}{\sqrt{2}}
\label{eq:hbasis}
\end{equation}
and $\rvh$, $\evh_\theta$, and $\evh_\phi$ are orthonormal vectors in spherical coordinates 
$(r,\theta,\phi)$. 
The orthogonality relations between these vectors are $\ve_{\pm}\cdot\ve_{\pm}=\rvh\cdot\rvh=1$ 
and $\ve_+\cdot\ve_-=\ve_{\pm}\cdot\rvh=0$, where the inner product between two vectors $\vu$ 
and $\vv$ is defined as $\vu\cdot\vv\equiv u_i \ov{v}_i\equiv u_i v^i$. Unless otherwise stated, 
an overline will denote complex conjugation throughout Sec.~\ref{sec:helicity}. 

In this reference frame, we decompose the first derivatives as
\begin{equation}
\veta\equiv \eta^{(0)}\rvh  + \eta^{(+1)}\evh_+ + \eta^{(-1)}\ve_-\;.
\label{eq:vdecomp}
\end{equation}
Here, $\eta^{(0)}\equiv\veta\cdot\rvh$ and $\eta^{(\pm 1)}\equiv\veta\cdot\ve_{\pm}$ are the 
helicity-0 and -1 components. The correlation properties of $\eta^{(0)}$ and $\eta^{(\pm 1)}$ 
can be obtained by projecting out the scalar and vector parts of the correlation of the 
Cartesian components $\eta_i$ with the projection operator 
$P=\ve_+\otimes\ov{\ve}_+ +\ve_-\otimes\ov{\ve}_-$. The rule of thumb is that 
$\la\eta_1^{(s)}\ov{\eta}_2^{(s')}\ra=\la\eta_1^{(s)}\eta_2^{(-s')}\ra$, where 
$\eta_\alpha^{(s)}=\eta^{(s)}(\vx_\alpha)$, vanish unless $s-s'=0$. We find
\begin{align}
\label{eq:xvector}
\la \eta^{(0)}_1\eta^{(0)}_2\ra &= 
\frac{1}{3\sigma_1^2}\bigl(\xi_0^{(1)}-2\xi_2^{(1)}\bigr) \\
\la \eta^{(\pm 1)}_1\ov{\eta}^{(\pm 1)}_2\ra &= 
\frac{1}{3\sigma_1^2}\bigl(\xi_0^{(1)}+\xi_2^{(1)}\bigr) \nonumber \\
\la \eta^{(\pm 1)}_1\ov{\eta}^{(\mp 1)}_2\ra &= 0 \nonumber \;.
\end{align}
Here and henceforth, the subscripts ``1'' and ``2'' will denote variables  evaluated at position 
$\vx_1$ and $\vx_2$ for shorthand convenience. Similarly, the symmetric tensor $\zeta_{ij}$ can 
be decomposed into its trace and traceless components, 
\begin{align}
\label{eq:tdecomp}
\zeta_{ij} &\equiv -\frac{1}{3}u\,\delta_{ij} + \tilde{\zeta}_{ij} \\
\tilde{\zeta}_{ij} &= S_{ij}\zeta^{(S)}  + \sqrt{\frac{1}{3}}
\left(\zeta^{(V)}_i\rh_j+\zeta^{(V)}_j\rh_i\right)  +
\sqrt{\frac{2}{3}} \zeta_{ij}^{(T)} \nonumber \;.
\end{align}
The variables $u\equiv -\tr\zeta=-\zeta_i^i$ and $\zeta^{(S)}\equiv\zeta^{(0)}$ are the
longitudinal and transverse helicity-0 modes, $\zeta^{(V)}_i$ are the components of a transverse 
vector, $\vzeta^{(V)}\cdot\rvh=0$, whereas $\zeta_{ij}^{(T)}$ is a symmetric, traceless, 
transverse tensor, $\delta^{ij}\zeta_{ij}^{(T)}=\zeta_{ij}^{(T)}\rh^j=0$.  
Explicit expressions for these variables are
\begin{align}
\zeta^{(S)} &\equiv \frac{3}{2} S^{lm}\zeta_{lm} = \frac{1}{2}
\left(3\rh^l\rh^m-\delta^{lm}\right)\zeta_{lm} \\ \zeta^{(V)}_i
&\equiv \sqrt{3}V_i^{lm}\zeta_{lm} =
\sqrt{3}\left(\delta_i^l-\rh_i\rh^l\right)\rh^m\zeta_{lm} \\
\zeta_{ij}^{(T)} &\equiv \sqrt{\frac{3}{2}}T_{ij}^{lm}\zeta_{lm}  \\ 
& = \sqrt{\frac{3}{2}}\left(P_i^l P_j^m - \frac{1}{2} P_{ij} P^{lm}\right)
\zeta_{lm} \nonumber \;,
\end{align}
where $S_{ab}$, $V_a^{bc}$ and $T_{ab}^{cd}$ are the scalar, vector and tensor projections 
operators (see, e.g., \cite{1994FCPh...15..209D}).
We have introduced factors of $\sqrt{1/3}$ and $\sqrt{2/3}$ in the decomposition 
Eq.(\ref{eq:tdecomp}) such that the zero-point moments of the helicity-0, -1 and -2 variables 
all equal $1/5$ (see Eq.~(\ref{eq:xtraceless}) below).
The helicity-1 components of $\vzeta^{(V)}$ and their complex conjugates are given by 
$\zeta^{(\pm 1)}\equiv \vzeta^{(V)}\cdot\ve_{\pm}=\sqrt{3}\,e_{\pm}^i\rh^j\zeta_{ij}$ and
$\czeta^{(\pm 1)}\equiv\vzeta^{(V)}\cdot\ov{\ve}_{\pm}=\sqrt{3}\,\ov{e}_{\pm}^i\rh^j\zeta_{ij}$, 
whereas $\zeta^{(\pm 2)}\equiv \zeta_{ij}^{(T)} e_{\pm}^i
e_{\pm}^j=\sqrt{3/2}\,e_{\pm}^i e_{\pm}^j\zeta_{ij}$ and  $\czeta^{(\pm
2)}\equiv \zeta_{ij}^{(T)} \ov{e}_{\pm}^i
\ov{e}_{\pm}^j=\sqrt{3/2}\,\ov{e}_{\pm}^i\ov{e}_{\pm}^j\zeta_{ij}$ are the two independent 
helicity-2 modes (polarizations) and their complex conjugates, respectively. Hereafter 
designating $\zeta^{(s)}(\vx_\alpha)$ as $\zeta_\alpha^{(s)}$, the correlation properties of 
these variables are the following:
\begin{align}
\label{eq:xtraceless}
\la \zeta_1^{(0)}\zeta_2^{(0)}\ra &= 
\frac{1}{\sigma_2^2} \left(\frac{1}{5}\xi_0^{(2)}
-\frac{2}{7}\xi_2^{(2)}+\frac{18}{35}\xi_4^{(2)}\right) \\
\la\zeta_1^{(\pm 1)}\czeta_2^{(\pm 1)}\ra &=
\frac{1}{\sigma_2^2}\left(\frac{1}{5}\xi_0^{(2)}
-\frac{1}{7}\xi_2^{(2)}-\frac{12}{35}\xi_4^{(2)}\right) \nonumber \\
\la\zeta_1^{(\pm 2)}\czeta_2^{(\pm 2)}\ra &=
\frac{1}{\sigma_2^2}\left(\frac{1}{5}\xi_0^{(2)}
+\frac{2}{7}\xi_2^{(2)}+\frac{3}{35}\xi_4^{(2)}\right) \nonumber \;,
\end{align}
and $\la\zeta_1^{(s)}\zeta_2^{(s')}\ra=\la\zeta_1^{(s)}\czeta_2^{(-s')}\ra$.
All the other correlations vanish. Note that the covariances are real despite the fact that 
the helicity-1 and -2 variables are complex.  

While the average peak number density only depends on the matrix $\vmm$ of covariances at 
the same location, the computation of the peak 2-point correlation function $\xpk(r)$  and 
higher-order clustering statistics from Eq.~(\ref{eq:Nptpk}) generally involve covariances of 
the random fields at different locations. 
For $\xpk(r)$, the covariance matrix $\vcc(r)\equiv\la\vy\vy^\dagger\ra$, where $\vy=(\vy_1,\vy_2)$, 
is a 20-dimensional matrix that may be partitioned into four $10\times 10$ block matrices: the 
zero-point contribution $\vmm$ in the top left and bottom right corners, and the cross-correlation 
matrix $\vbb(r)$ and its transpose in the bottom left and top right corners, respectively. 
Expressions for $\vmm$ and $\vbb(r)$ in the helicity basis can be found in the Appendix of 
\cite{2010PhRvD..82j3529D}.

\subsection{Rotational invariants}

Translational and rotational invariance implies that $\bnpk$ does not depend on spatial position, 
and that $\xpk(r)$ be a function of the distance $r$ solely. 
In this regards, \cite{2010PhRvD..82j3529D} noted that, although the covariance matrix $\vcc(r)$
in the helicity basis (\ref{eq:hbasis}) does not depend on the direction $\rvh$ of the separation 
vector $\vr$, it is not equal to the angular average covariance matrix 
$\wi{\vcc}(r)\equiv(1/4\pi)\int\!d\Omega_{\rvh}\,\vcc(\vr)$. 
The latter is obtained upon setting $\xi_\ell^{(n)}\equiv 0$ whenever $\ell\neq 0$ in the expression 
of $\vbb(r)$. As a consequence, $\wi{\vcc}(r)$ retains the correlations $\la\nu_1\nu_2\ra$, 
$\la\nu_1 u_2\ra$, $\la u_1 u_2\ra$ and parts of the covariances $\la\eta_1^{(m)}\ov{\eta}_2^{(m)}\ra$ 
and $\la\zeta_1^{(m)}\czeta_2^{(m)}\ra$. However, \cite{2010PhRvD..82j3529D} did not provide a 
convincing explanation for their observation. We shall do it now.

Firstly, it is pretty clear that, since the peak 2-point correlation is invariant under rotations 
of the reference frame, it should be possible to express it in terms of rotational invariants 
constructed from the variables $\nu$, $\eta_i$ and $\zeta_{ij}$. 
The peak significance $\nu$ and the trace $u$ are two obvious candidates, but they are not the 
only ones. The vector $\veta$ of first derivatives and the traceless matrix $\tilde{\zeta}$ yield 
two additional invariants, i.e. the square modulus $\veta\cdot\veta$ and the trace 
$\tr(\tilde{\zeta}^2)$. 
In the helicity basis, these invariants can be written
\begin{equation}
\label{eq:modulus}
\veta\cdot\veta = \eta^{(0)}\eta^{(0)}
+\eta^{(+1)}\ov{\eta}^{(+1)}+\eta^{(-1)}\ov{\eta}^{(-1)}
\end{equation}
and
\begin{multline}
\label{eq:trace2}
\tr\bigl(\tilde{\zeta}^2\bigr) = \frac{2}{3} \biggl[
\zeta^{(0)}\zeta^{(0)} \\ +\sum_{s=1,2}\Bigl(\zeta^{(+s)}\czeta^{(+s)}
+ \zeta^{(-s)}\czeta^{(-s)}\Bigr) \biggr] \;.
\end{multline}
The $3\times 3$ symmetric matrix $\zeta_{ij}$ actually provides a third invariant with respect to 
rotations: 
the determinant det$\zeta$. However, as we shall see below in the discussion of the peak-background 
split, because this determinant only enters the peak number density $\npk(\vy)$ and not the 1-point 
multivariate normal distribution $P_1(\vy)$ (the argument of the exponential is quadratic in the 
variables), it does not contribute directly to the peak bias.
This suggests that we look at the covariances of $\eta^2(\vx)$ and $\zeta^2(\vx)$, where these 
are defined as
\begin{widetext}
\begin{align}
\eta^2(\vx) &\equiv \veta(\vx)\cdot\veta(\vx) =
-\frac{1}{\sigma_1^2}\int\!\!\frac{d^3k_1}{(2\pi)^3}\int\!\!\frac{d^3k_2}{(2\pi)^3}\,
\delta_{ij} k_{1i} k_{2j}\, \delta_s(\vk_1)\delta_s(\vk_2) e^{i(\vk_1+\vk_2)\cdot\vx}\;, \\
\zeta^2(\vx) &\equiv \frac{3}{2}\tr\Bigl[\tilde{\zeta}^2(\vx)\Bigr] =
\frac{3}{2\sigma_2^2}\int\!\!\frac{d^3k_1}{(2\pi)^3}\int\!\!\frac{d^3k_2}{(2\pi)^3}\,
\left(\delta_{il}\delta_{jm}-\frac{1}{3}\delta_{ij}\delta_{lm}\right)
k_{1i} k_{1j} k_{2l} k_{2m}\,\delta_s(\vk_1)\delta_s(\vk_2) e^{i(\vk_1+\vk_2)\cdot\vx} \;,
\end{align}
\end{widetext}
where $\delta_s(\vk)$ are the Fourier modes of the smoothed density field.
As we shall see in Sec.~\ref{sec:pbs}, the variables $3\eta^2$ and $5\zeta^2$ are distributed as 
chi-squared ($\chi^2$) variates with 3 and 5 degrees of freedom, respectively.
Using either the Fourier space expression of $\eta_\alpha^2\equiv\eta^2(\vx_\alpha)$ (not to be 
confonded here with a cartesian component of the vector $\veta$) or the fact that only 
components of identical helicity correlate, it is straightforward to compute the following 
correlators (for illustrative purposes, Appendix \ref{sec:correlators} furnishes details of the 
calculation of $\la\eta_1^2\eta_2^2\ra$):
\begin{align}
\label{eq:eta1eta2}
\la\eta_1^2\eta_2^2\ra &= 1 + \frac{2}{3\sigma_1^4}
\Bigl[\bigl(\xi_0^{(1)}\bigr)^2+ 2\bigl(\xi_2^{(1)}\bigr)^2\Bigr] \\
\la\eta_1^2\nu_2^2\ra &= 1 + \frac{2}{\sigma_0^2\sigma_1^2}\,\xi_1^{(1/2)}\xi_1^{(1/2)} \;.
\end{align}
Ignoring the zero-lag contributions, these terms correspond exactly (up to a sign factor) to some 
of those entering the second-order contribution of $\xpk(r)$ in Eq.(\ref{eq:xpkeasy}), with 
$(\xi_1^{(1/2)}b_{20}\xi_1^{(1/2)})$ being proportional to $\la\eta_1^2\nu_2^2\ra$ in particular. 
The computations of correlators involving $\zeta_\alpha^2\equiv \zeta^2(\vx_\alpha)$ proceeds in a 
similar way although, in this case, it is much easier to sum the correlations among equal helicity 
components. For instance, 
\begin{align}
\la\zeta_1^2\zeta_2^2\ra &= 1+ 2\,\biggl\{\la\zeta_1^{(0)}\zeta_2^{(0)}\ra^2 \\
&\quad +2 \sum_{s=1,2}\Bigl(\la\zeta_1^{(+s)}\czeta_2^{(+s)}\ra^2
+\la\zeta_1^{(-s)}\czeta_2^{(-s)}\ra^2\Bigr)\biggr\} \nonumber \;,
\end{align}
where we used the fact that $\la\zeta_1^{(s)}\zeta_2^{(-s)}\ra=\la\zeta_1^{(s)}\czeta_2^{(s)}\ra$.
After some algebra, we find
\begin{multline}
\la\zeta_1^2\zeta_2^2\ra = 1+ \frac{2}{5\sigma_2^4}\Bigl[\bigl(\xi_0^{(2)}\bigr)^2 \\ 
+\frac{10}{7} \bigl(\xi_2^{(2)}\bigr)^2+\frac{18}{7}\bigl(\xi_4^{(2)}\bigr)^2\Bigr]\;,
\end{multline}
and
\begin{align}
\la\zeta_1^2\eta_2^2\ra &= 1 + \frac{2}{5\sigma_1^2\sigma_2^2}
\Bigl[3\bigl(\xi_3^{(3/2)}\bigr)^2+2\bigl(\xi_1^{(3/2)}\bigr)^2\Bigr] \\
\la\zeta_1^2\nu_2^2\ra &= 1 + \frac{2}{\sigma_0^2\sigma_2^2}\xi_2^{(1)} \xi_2^{(1)} \;.
\end{align}
The cross-correlations of $\eta_1^2$ and $\zeta_1^2$ with $u_2$ in place of $\nu_2$ are identical 
except for the superscript $(n)$, which should be replaced by $(n+1)$.
Again, all these terms can be found among the second-order contributions in the right-hand side 
of Eq.~(\ref{eq:xpkeasy}).

Therefore, the actual dependence of $\xpk(r)$ on the invariants $\eta^2(\vx)$ and $\zeta^2(\vx)$, 
whose covariances involve the correlation functions $\xi_\ell^{(n)}$ with $\ell\neq 0$, is the 
fundamental reason for $\vcc(r)$ being different from the angle average $\wi{\vcc}(r)$. Those 
correlations, which arise upon expanding the joint probability density at second order, eventually 
all nicely combine together (and with terms proportional to $(\xi_0^{(1)})^2$ and $(\xi_0^{(2)})^2$) 
to yield the second-order correlators $\la\eta_1^2\nu_2^2\ra$, $\la\zeta_1^2\eta_2^2\ra$ etc.

The question then arises of the calculation of the coefficients of these quadratic terms in the 
peak 2-point correlation function. We already know that the coefficients multiplying products 
of the form $\xi_0^{(n)}\xi_0^{(n')}$ are the quadratic peak-background split biases associated 
to the scalars $\nu$ and $u$. Does this hold also for the coefficients multiplying 
$\la\eta_1^2\eta_2^2\ra$, $\la\zeta_1^2\nu_2^2\ra$ etc. ?

\subsection{Generalizing the peak background-split}
\label{sec:pbs}

The probability distribution $P_1(\vy)$ that is needed to compute $\bnpk$ is a multivariate 
Gaussian of covariance matrix $\vmm$, 
\begin{equation}
P_1(\vy) d^{10}\vy= \frac{1}{(2\pi)^5|\det\vmm|^{1/2}} e^{-Q_1(\vy)} d^{10}\vy \;.
\end{equation}
Owing to rotational invariance, this probability density is a function of $\nu$, $u$, $\eta^2$ 
and $\zeta^2$ solely (see, e.g., \cite{2009PhRvD..80h1301P} for a systematic analysis of 
distribution functions of homogeneous and isotropic random fields). 
The quadratic form $Q_1(\vy)$ that appears in the exponential factor reads
\begin{equation}
Q_1(\vy) = 
\frac{\nu^2+u^2-2\gamma_1\nu u}{2\left(1-\gamma_1^2\right)}
+\frac{3}{2}\,\eta^2 + \frac{5}{2}\,\zeta^2 \;,
\end{equation}
so that $\exp[-Q_1(\vy)]$ retains factorization with respect to $(\nu,u)$, $\eta^2$ and 
$\zeta^2$. Furthermore, since $\eta^{(s)}$ (with $s=\pm 1$) and $\zeta^{(s)}$ (with 
$s=\pm 1,\pm 2$) are complex random variables with mean 0 and variance $1/3$ and $1/5$
respectively and since 
$\la \nu \zeta^{(0)}\ra =\la \nu\eta^{(0)}\ra = 0$ and $\la\eta^{(0)}\zeta^{(0)}\ra=0$ at 
the same spatial location, the quantities $3\eta^2(\vx)$ and $5\zeta^2(\vx)$ are 
independent $\chi^2$-distributed variables with 3 and 5 degrees of freedom, respectively 
(similar conclusions can be drawn for the distribution of the components of the deformation 
tensor, see \cite{2002MNRAS.329...61S,2012arXiv1207.7117S}). 
Therefore, the 1-point probability density can also be written as
\begin{multline}
\label{eq:P1}
P_1(\vy) d^{10}\vy = {\cal N}(\nu,u)\,d\nu du \\ \times 
\chi_3^2(3\eta^2)\, d(3\eta^2)\, \chi_5^2(5\zeta^2)d(5\zeta^2)\;,
\end{multline}
where ${\cal N}(\nu,u)$ is the bivariate normal
\begin{equation}
  {\cal N}(\nu,u) = \frac{1}{2\pi\sqrt{1-\gamma_1^2}}
\exp\left[-\frac{\nu^2+u^2-2\gamma_1\nu u}{2(1-\gamma_1^2)}\right] \;,
\end{equation}
and 
\begin{equation}
\chi_k^2(x) = \frac{1}{2^{k/2}\Gamma(k/2)} x^{k/2-1} e^{-x/2}
\end{equation}
is a $\chi^2$-distribution with $k$ degrees of freedom. Note that the distribution of 
$\zeta^2$ is coupled with the last rotational invariant $\det\zeta$ \cite{2012PhRvD..85b3011G}.
However, it can be easily checked that, upon integrating over the (uniform) distribution of 
$\det\zeta$, we obtain the $\chi^2$-distribution $\chi^2_5(5\zeta^2)$.

Ref.~\cite{2010PhRvD..82j3529D} discussed how the peak bias factors $b_{ij}$ can be 
derived from a peak-background split. They argued that, while the $k$-dependent piece $b_{n0}$
is related to the $n$th order derivative of the differential number density $\bnpk$, 
derivatives cannot produce the bias factors $b_{01}$, $b_{11}$ etc. multiplying the 
$k$-dependent terms. For this reason, they considered a second implementation of the 
peak-background split \cite{1996MNRAS.282..347M,1999MNRAS.308..119S} in which the dependence 
of the mass function on the overdensity of the background is derived explicitly. However, it 
is possible to write all the peak bias factors $b_{ij}$ as derivatives of ${\cal N}(\nu,u)$ 
rather than $\bnpk$. More precisely, the $b_{ij}$ are the bivariate Hermite polynomials
\begin{equation}
H_{ij}(\nu,u)={\cal N}(\nu,u)^{-1}
\left(-\frac{\partial}{\partial\nu}\right)^i 
\left(-\frac{\partial}{\partial u}\right)^j {\cal N}(\nu,u)
\end{equation}
relative to the weight ${\cal N}(\nu,u)$, further averaged over the peak curvature $u$. 
Therefore, they are peak-background split biases in the sense that they can be derived from 
the transformation $\nu\to\nu+\epsilon_1$ and $u\to u+\epsilon_2$, where $\epsilon_i$ and 
$\epsilon_2$ are long-wavelength background perturbations uncorrelated with the (small-scale)
fields $\nu(\vx)$ and $u(\vx)$. 
Ref.~\cite{2010PhRvD..82j3529D} did not express the peak-background split this way 
because they considered the effect of a background perturbation {\it after} the integration 
over the peak curvature. 
In terms of the rotational invariants introduced above, we can write the $b_{ij}$ as
\begin{equation}
\sigma_0^i \sigma_2^j b_{ij}=\frac{1}{\bnpk}\int\!\!d^{10}\vy\,\npk(\vy) 
H_{ij}(\nu,u) P_1(\vy) \;,
\end{equation}
where it is understood that $P_1(\vy)$ takes the form Eq.~(\ref{eq:P1}) and 
$d^{10}\vy=d\nu\, du\, d(3\eta^2)\, d(5\zeta^2)$. Factors of $1/\sigma_0$ and $1/\sigma_2$ are
introduced because bias factors are ordinarily defined relative to the physical field 
$\delta_s(\vx)$ and its derivatives rather than the normalized variables.
In practice, the integral is most easily performed
upon transforming the 5 degrees of freedom attached to $5\zeta^2$ to the shape parameters $v$ 
and $w$ and the 3 Euler angles that describe the orientation of the principal axis frame
(see Appendix \ref{sec:uvw} for details).

In \cite{2012PhRvD..85b3011G}, it was noticed that, in the presence of non-Gaussianity, the 
1-point probability density $P_1(\vy)$ can be expanded in the set of orthogonal polynomials 
associated to the weight provided by $P_1(\vy)$ in the Gaussian limit. The same logic applies
to the peak bias factors. Namely, the $b_{ij}$ are drawn from the orthogonal polynomials 
associated to ${\cal N}(\nu,u)$, i.e. bivariate Hermite polynomials. Therefore, we expect that
$\eta^2(\vx)$ and $\zeta^2(\vx)$ also generate bias parameters, and that these are drawn from
the orthogonal polynomials associated with $\chi^2$-distributions, i.e. generalized Laguerre
polynomials. These are defined as
\begin{equation}
\label{eq:laguerre}
L_n^{(\alpha)}(x) = 
\frac{x^{-\alpha} e^x}{n!} \frac{d^n}{dx^n}\!\left(e^{-x} x^{n+\alpha}\right)
\end{equation}
and are orthogonal over $[0,\infty[$ with respect to the $\chi^2$-distribution with 
$k=2(\alpha+1)$ degrees of freedom. The orthogonality relation can be expressed as
\begin{equation}
\label{eq:ortho}
\int_0^\infty\!\!dx\,x^\alpha e^{-x} L_n^{(\alpha)}(x) L_m^{(\alpha)}(x)
= \frac{\Gamma(n+\alpha+1)}{n!}\, \delta_{mn} \;.
\end{equation}
The first generalized Laguerre polynomials are 
$L_0^{(\alpha)}(x)=1$ and $L_1^{(\alpha)}(x) = -x+\alpha + 1$.

Given the correlator Eq.~(\ref{eq:eta1eta2}), the term proportional to $1/\sigma_1^4$ in the 
right-hand side of Eq.(\ref{eq:xpkeasy}) indicates that the first-order bias parameter 
$\chi_{10}$ associated with the invariant $\sigma_1^2\eta^2(\vx)=(\grad\delta_s)^2(\vx)$ is 
$\chi_{10}=-3/(2\sigma_1^2)$. The aforementioned considerations suggest that we define the 
$k$th-order bias factor as the Laguerre polynomial $(-1)^k L_k^{(1/2)}(3\eta^2/2)$ averaged 
over all the possible peak configurations, i.e.
\begin{equation}
\sigma_1^{2k} \chi_{k0} \equiv \frac{(-1)^k}{\bnpk}\int\!\!d^{10}\vy\,\npk(\vy)
L_k^{(1/2)}\!\!\left(\frac{3\eta^2}{2}\right)\, P_1(\vy)\;.
\end{equation}
Taking into account the peak constraint, the first-order bias factors thus is
\begin{align}
\chi_{10} &= \frac{1}{\sigma_1^2\bnpk}
\int\!\!d^{10}\vy\,\npk(\vy) \left(\frac{3}{2}\eta^2-\frac{3}{2}\right) P_1(\vy) \\
&= -\frac{3}{2\sigma_1^2} \;,
\end{align}
which is precisely what we were expecting. Similarly, we define the bias parameter $\chi_{0k}$ 
associated to the invariant $\zeta^2(\vx)$ as the ensemble average of the Laguerre polynomial 
$L_k^{(3/2)}(5\zeta^2/2)$ orthogonal with respect to the weight $\chi_5^2(5\zeta^2)$. Namely,
\begin{equation}
\sigma_2^{2k} \chi_{0k} \equiv \frac{(-1)^k}{\bnpk}\int\!\!d^{10}\vy\,\npk(\vy)
L_k^{(3/2)}\!\!\left(\frac{5\zeta^2}{2}\right)\, P_1(\vy) \\
\end{equation}
Note that, although we use a single symbol $\chi_{ij}$ to designate the bias factors derived 
from the $\chi^2$-distributions, the variables $\eta^2$ and $\zeta^2$, unlike $\nu$ and $u$, 
are uncorrelated. The first-order bias factor thus is 
\begin{equation}
\chi_{01} = \frac{1}{\sigma_2^2\bnpk}
\int\!\!d^{10}\vy\,\npk(\vy) \left(\frac{5}{2}\zeta^2-\frac{5}{2}\right)P_1(\vy) \;.
\end{equation}
To evaluate the integral, we first express the measure $d(5\zeta^2)$ in terms of the ellipticity 
$v$ and prolateness $w$, so that $\zeta^2$ can be written as $\zeta^2=3v^2+w^2$ (see Appendix 
\ref{sec:uvw} for details). 
A multiplicative factor of $\zeta^2$ will arise upon, e.g., taking the derivative of 
$\exp(-5\alpha\zeta^2/2)$ with respect to $\alpha$. 
In the notation of \cite{2010PhRvD..82j3529D}, our factor of $\zeta^2$ precisely corresponds to 
their derivative term $-(2/5)\partial_\alpha\ln G_0^{(\alpha)}\!(\gamma_1,\gamma_1\nu)$ evaluated 
at $\alpha=1$ (see Appendix \ref{sec:uvw}). Taking into account the factor of 
$G_0^{(1)}\!(\gamma_1,\gamma_1\nu)$ in the denominator, $\chi_{01}$ can eventually be written
\begin{equation}
\label{eq:bzz}
\chi_{01}=-\frac{5}{2\sigma_2^2} \left(1+\frac{2}{5}
\partial_\alpha\ln G_0^{(\alpha)}\!(\gamma_1,\gamma_1\nu)\Bigr\rvert_{\alpha=1}\right)\;.
\end{equation}
The physical interpretation of this result is straightforward: $\zeta^2(\vx)$ is a scalar that 
describes the asymmetry of the peak density profile. In the high peak limit, $\chi_{01}\to 0$ 
reflecting the fact that the most prominent peaks are nearly spherical 
(see Fig.9 of \cite{2010PhRvD..82j3529D}). 

The physical origin for the appearance of these orthogonal polynomials can be found in the 
peak-background split. 
Long-wavelength background perturbations locally modulate the mean of the distributions 
${\cal N}(\nu,u)$, $\chi_3^2(3\eta^2)$ and $\chi_5^2(5\zeta^2)$. The resulting non-central 
distributions can then be expanded in the appropriate set of orthogonal polynomials. 
In practice, it is convenient to introduce a shift or translation operator $\hat{T}_\epsilon$ 
to describe the action of a background perturbation on the distribution of rotational invariants. 
For the scalars $\nu$ and $u$, we define the shift operator as
\begin{equation}
\hat{T}_\epsilon \equiv \exp\left(-\epsilon_1\partial_\nu -\epsilon_2\partial_u\right) \;,
\end{equation}
where $\epsilon_1$ and $\epsilon_2$ are small perturbations to the peak significance and the 
peak curvature, i.e. $\nu\to\nu+\epsilon_1$ and $u\to u+\epsilon_2$.
The action of $\hat{T}_\epsilon$ on the probability density ${\cal N}(\nu,u)$ is to shift the
(zero) mean of $\nu$ and $u$ by $-\epsilon_1$ and $-\epsilon_2$, respectively (the reason for
the minus sign is that Hermite polynomials include a factor of $(-1)^{i+j}$). A straightforward 
calculation gives
\begin{align}
\label{eq:Tnormal}
{\cal N}(\nu,u)^{-1}\, \hat{T}_\epsilon\, {\cal N}(\nu,u) &= 
\frac{{\cal N}(\nu-\epsilon_1,u-\epsilon_2)}{{\cal N}(\nu,u)} \\
&= f(\epsilon_1,\epsilon_2)\, e^{\epsilon_1\sigma_0b_\nu +\epsilon_2\sigma_2 b_u} 
\nonumber \;,
\end{align}
where $b_\nu$ and $b_u$ are given in Eqs.(\ref{eq:bv}) and (\ref{eq:bu}), and 
$f(\epsilon_1,\epsilon_2)$ is the exponential factor in ${\cal N}(\nu,u)$ with the replacement 
$\nu\to\epsilon_1$ and $u\to\epsilon_2$. The last expression
is a generating function of bivariate Hermite polynomials. On expanding it in the small 
parameters $\epsilon_1$ and $\epsilon_2$, 
\begin{multline}
\Bigl\la f(\epsilon_1,\epsilon_2)\, e^{\epsilon_1\sigma_0 b_\nu +\epsilon_2\sigma_2 b_u}
\Bigr\lvert{\rm pk}\Bigr\ra \\
= \sum_{i,j=0}^\infty \sigma_0^i\sigma_2^j b_{ij}
\biggl(\frac{\epsilon_1^i}{i!}\biggr)
\biggl(\frac{\epsilon_2^j}{j!}\biggr)\;,
\end{multline}
we recover the bias factors $b_{ij}$ once the results are averaged over all locations that 
satisfy the peak constraint. Note that the bias parameter $b_{\nabla^2\delta}$ defined in
\cite{2012arXiv1212.0868S} bears the same physical meaning as our $b_{01}$: both represent 
the leading-order response of the tracer abundance to a uniform shift in the curvature of 
the density field.

For the quadratic variables $\eta^2$ and $\zeta^2$, Eq.(\ref{eq:laguerre}) suggests that we 
express the shift operator in terms of both $x$ and $\partial_x$. The definition is somewhat
cumbersome because we must take into account not only the ordering of $x$ and $\partial_x$,
but also the factor of $\Gamma(n+\alpha+1)/n!$ in the orthogonality relation Eq.(\ref{eq:ortho}).
A sensible definition of $\hat{T}_\epsilon$ for the variable $x=3\eta^2$ and $5\zeta^2$ is
\begin{align}
\hat{T}_\epsilon &\equiv\, 
:\sum_{j=0}^\infty \frac{\Gamma\bigl(\frac{1}{2}k\bigr)}
{j!\,\Gamma\bigl(\frac{1}{2}k+j\bigr)}\left(-\frac{\epsilon}{2}\partial_x x\right)^j: \\
&= \Gamma\bigl(\frac{1}{2}k\bigr) \,:
\frac{I_\alpha\!\left(\sqrt{-\frac{\epsilon}{2}\partial_x x}\right)}
{\left(-\frac{\epsilon}{2}\partial_x x\right)^{\alpha/2}}:\quad\;,
\end{align}
where $I_\alpha(x)$ is a modified Bessel function of the first kind and the symbol $::$ of 
normal ordering is borrowed from quantum field theory. In the present discussion, the normal 
ordering is defined as
\begin{equation}
:\!\left(\partial_x x\right)^n\!:\, f(x) \equiv \partial_x^n\bigl( x^n f(x) \bigr) \;,
\end{equation}
where $f(x)$ is some test function. With this definition, the action of $\hat{T}_\epsilon$ on
a $\chi^2$-distribution with $k=2(\alpha+1)$ degrees of freedom is
\begin{align}
\label{eq:Tchi}
\hat{T}_\epsilon\,& \chi_k^2(x) \\
&= \frac{1}{2} 
:\!\left[\sum_{j=0}^\infty \frac{\left(-\frac{\epsilon}{2}\right)^j}
{j!\,\Gamma\bigl(\frac{1}{2}k+j\bigr)}\left(\partial_x x\right)^j\right]\!:\,
\left(\frac{x}{2}\right)^\alpha e^{-x/2} \nonumber \\
&= \frac{1}{2}\sum_{j=0}^\infty \frac{\left(-\frac{\epsilon}{2}\right)^j}
{j!\,\Gamma\bigl(\frac{1}{2}k+j\bigr)}\partial_{\frac{x}{2}}^j
\left[\left(\frac{x}{2}\right)^{\alpha+j} e^{-x/2}\right] \nonumber \\
&= \frac{e^{-x/2}}{2}\left(\frac{x}{2}\right)^\alpha 
\sum_{j=0}^\infty\frac{\left(-\frac{\epsilon}{2}\right)^j}
{\Gamma\bigl(\frac{1}{2}k+j\bigr)} L_j^{(\alpha)}\!\left(\frac{x}{2}\right)
\nonumber \\
&= \chi_k^2(x)\sum_{j=0}^\infty \frac{\Gamma\bigl(\frac{1}{2}k\bigr)}
{\Gamma\bigl(\frac{1}{2}k+j\bigr)}\left(-\frac{\epsilon}{2}\right)^j
L_j^{(\alpha)}\!\left(\frac{x}{2}\right) \nonumber \;.
\end{align}
This is precisely the Laguerre series expansion of a non-central $\chi^2$-variate derived in 
\cite{tiku:1965} (see Appendix \ref{sec:chisquare}). Therefore, 
\begin{equation}
\left[\chi_k^2(x)\right]^{-1}\hat{T}_\epsilon\, \chi_k^2(x) 
= \frac{\chi_k^{'2}(x;\epsilon)}{\chi_k^2(x)} \nonumber \;,
\end{equation}
where $\chi_k^{'2}(x;\lambda)$ is a non-central $\chi^2$-distribution with $k$ degrees of 
freedom and non-centrality parameter $\lambda\equiv\epsilon$. The latter is defined as the 
sum of squares $\lambda=\sum_{i=1}^k \mu_i^2$, where $\mu_i$ are the means of the random 
variables.
We can now read off the bias factors from the expansion of 
$[\chi_k^2]^{-1} \hat{T}_\epsilon\chi_k^2$ in generalized Laguerre polynomials. For instance,
\begin{equation}
\biggl\la\frac{\chi_5^{'2}(5\zeta^2;\epsilon_5)}{\chi_k^2(5\zeta^2)}\biggr\lvert{\rm pk}\biggr\ra
= \sum_{j=0}^\infty \frac{\Gamma\bigl(\frac{5}{2}\bigr)}
{\Gamma\bigl(\frac{5}{2}+j\bigr)}\left(\frac{\epsilon_5}{2}\right)^j\sigma_2^{2j}\chi_{0j} \;. 
\end{equation}
Note that the more common generating function
\begin{equation}
\left(1-\epsilon\right)^{-\alpha-1} 
\exp\!\Biggl[\frac{x\epsilon}{2\bigl(1-\epsilon\bigr)}\Biggr] 
= \sum_{n=0}^\infty \epsilon^n L_n^{(\alpha)}\!(x)
\end{equation}
appears to bear little connection with the non-central $\chi^2$-distribution.

To better understand the reason why the peak-background split generates a non-central 
$\chi^2$-distribution, we note that, owing to the relation $H_{2k}(x)\sim L_k(x^2)$ between 
Hermite and Laguerre polynomials, we could also have defined $\chi_{10}$ and $\chi_{01}$ as 
second derivatives of $P_1(\vy)$, where $\vy$ is now the vector $(\nu,u,\eta_i,\tilde{\zeta}_A)$ 
of independent normal random variables such that $\eta^2=\sum_{i=1}^3 \eta_i^3$ and 
$\zeta^2\equiv \sum_{A=1}^5\tilde{\zeta}_A^2$. A little algebra shows that
\begin{equation}
\chi_{10} \equiv \frac{1}{\bnpk}\int\!\!d^{10}\vy\,\npk(\vy)\sum_{j=1}^3
\left(\frac{1}{\sigma_1}\frac{\partial}{\partial\eta_j}\right)^2 P_1(\vy)\;,
\end{equation}
Another way of writing this formula would be to absorb a factor of $1/\sqrt{2}$ in the 
definition of $\eta^{(\pm 1)}$ (and thus explicitly deal with complex normal distributions). 
Analogously, we have
\begin{equation}
\chi_{01}\equiv \frac{1}{\bnpk}\int\!\!d^{10}\vy\,\npk(\vy) \sum_{A=1}^5
\left(\frac{1}{\sigma_2}\frac{\partial}{\partial\tilde{\zeta}_A}\right)^2 P_1(\vy)\;.
\end{equation}
This suggests that the effect of a background perturbation on $\eta^2$ and $\zeta^2$ can also
be thought of as shifting the components of the first derivatives according to 
$\eta_i\to\eta_i+\epsilon_{3i}$, and those of $\tilde{\zeta}$ according to 
$\tilde{\zeta}_A\to\tilde{\zeta}_A+\epsilon_{5A}$. The small perturbations $\epsilon_{3i}$ and
$\epsilon_{5A}$ need not be the same for distinct $i$ and/or $A$. However, owing to invariance
under rotations, only the length of the vector $\sum_{i=1}^3(\epsilon_{3i})^2\equiv\epsilon_3$
and $\sum_{A=1}^5(\epsilon_{5A})^2\equiv \epsilon_5$ matter. This is the reason why the 
background perturbation effectively shifts the respective $\chi^2$-distributions to non-central
$\chi^2$-distributions, with non-centrality parameter $\lambda=\epsilon_3$ and 
$\lambda=\epsilon_5$. Note that it should be possible to formulate this peak-background split 
with the conditional peak number density $\bnpk(\nu,R_s|\delta_l,R_l)$ in a large-scale region of 
overdensity $\delta_l$, like in \cite{2010PhRvD..82j3529D}. However, one should then consider 
two long-wavelength perturbations $\eta_l^2$ and $\zeta_l^2$ in order to describe the effect of 
the background perturbation on $\eta^2(\vx)$ and $\zeta^2(\vx)$, in addition to $\delta_l$ 
(which suffices to describe the effect of the background wave for both $\nu$ and $u$ since these 
variables are correlated).

To conclude, \cite{2011PhRvD..83h3518M} also pointed out that, even though there is no functional 
relation $\npk={\cal F}(\delta)$ for discrete density peaks, it is nevertheless possible to 
define renormalized bias parameters as the expectation values $c_n\sim\la{\cal F}\ra$. 
However, he did not compute them explicitly, nor specified what is ${\cal F}$ (though it is pretty 
clear that it is related to $\npk$). Here, we demonstrated explicitly that each of the combinations 
$(\nu,u)$, $\eta^2$ and $\zeta^2$ of rotational invariants generates a set of orthogonal polynomials
which, upon taking the ensemble average over all the possible peak configurations, yields a set 
of bias factors.
Furthermore, we showed that these bias parameters can be constructed from a suitable application 
of the peak-background split to the probability densities characterizing the invariants. 
We will now demonstrate that we can interpret the peak 2-point correlation as arising from a 
functional relation of the form $\dpk={\cal F}(\delta_s,\dots)$.

%two ways: i) the terms
%linear in $\bias{II}$ should have a plus sign rather than a minus sign and ii) the last term in the
%right-hand side of Eq.(\ref{eq:xpkeasy}) misses a multiplicative factor of 

%we could find the sign mistake in Eq.(A35), where $Q$ is a matrix that is quadratic in the correlation 
%$\xi_\ell^{(n)}(r)$.

\subsection{A local bias approach to $\xpk(r)$}
\label{sec:local}

First, let us make sure that \cite{2010PhRvD..82j3529D} obtained the correct expression for $\xpk(r)$. 
Adding all the second-order contributions induced by $\eta^2$, $\zeta^2$ and their cross-correlations 
with  $\nu$ and $u$, our result differs from theirs in that the last term in the right-hand side of 
Eq.(\ref{eq:xpkeasy}) appears to miss a multiplicative factor of 
$1+(2/5)\partial_\alpha\ln G_0^{(\alpha)}\!(\gamma_1,\gamma_1\nu)|_{\alpha=1}$.
Checking the calculation of \cite{2010PhRvD..82j3529D}, we found the missing multiplicative factor in 
their Eqs.~(A50) and (A51), in the form of $q(r)\tr(\tilde{\zeta}_i^2)$. This term was fortuitously 
omitted in their final expression of $\xpk(r)$. Consequently, the correct answer is 
\begin{widetext}
\begin{align}
\label{eq:xpknew}
\xpk(\nu,R_s,r) &= \bigl(\bias{I}^2\xi_0^{(0)}\bigr) +\frac{1}{2}
\bigl(\xi_0^{(0)}\bias{II}^2\xi_0^{(0)}\bigr)-\frac{3}{\sigma_1^2}
\bigl(\xi_1^{(1/2)}\bias{II}\xi_1^{(1/2)}\bigr)
-\frac{5}{\sigma_2^2}\bigl(\xi_2^{(1)}\bias{II}\xi_2^{(1)}\bigr)
\biggl(1+\frac{2}{5}\partial_\alpha\ln G_0^{(\alpha)}\!(\gamma_1,\gamma_1\nu)
\Bigl\rvert_{\alpha=1}\biggr) \nonumber \\
&\quad + \frac{5}{2\sigma_2^4}\Bigl[\bigl(\xi_0^{(2)}\bigr)^2+\frac{10}{7}
\bigl(\xi_2^{(2)}\bigr)^2+\frac{18}{7}\bigl(\xi_4^{(2)}\bigr)^2\Bigr]
\biggl(1+\frac{2}{5}\partial_\alpha\ln G_0^{(\alpha)}\!(\gamma_1,\gamma_1\nu)
\Bigl\rvert_{\alpha=1}\biggr)^2 \\ 
&\quad +\frac{3}{2\sigma_1^4}\Bigl[\bigl(\xi_0^{(1)}\bigr)^2
+2\bigl(\xi_2^{(1)}\bigr)^2\Bigr]+\frac{3}{\sigma_1^2\sigma_2^2}
\Bigl[3\bigl(\xi_3^{(3/2)}\bigr)^2+2\bigl(\xi_1^{(3/2)}\bigr)^2\Bigr]
\biggl(1+\frac{2}{5}\partial_\alpha\ln G_0^{(\alpha)}\!(\gamma_1,\gamma_1\nu)
\Bigl\rvert_{\alpha=1}\biggr)
\nonumber\;.
\end{align}
\end{widetext}
We note that this omission has an impact only on the small-scale ($r\lesssim 20\hmpc$) peak 
correlation displayed in Fig.2 of their paper. Their results concerning the peak-background split,
the gravitational evolution of $\xpk(r)$ or the scale-dependence of bias around the Baryon 
Acoustic Oscillation are unaffected.

Even though we cannot write down a relation of the form $\npk={\cal F}_X(\delta)$, the 
peak correlation function up to second-order can nonetheless be thought of as arising 
from a local bias expansion $\dpk={\cal F}_X(\delta_s,\dots)$, i.e. 
\begin{align}
\label{eq:newdpk}
\dpk(\vx) &= b_{10} \delta_s(\vx) - b_{01} \nabla^2\delta_s(\vx) 
+ \frac{1}{2} b_{20}\delta_s^2(\vx) \\ 
& -  b_{11}\delta_s(\vx)\nabla^2\delta_s(\vx) 
+ \frac{1}{2} b_{02}\bigl[\nabla^2\delta_s(\vx)\bigr]^2 \nonumber \\
& + \chi_{10}\left(\nabla\delta_s\right)^2\!\!(\vx) + \frac{1}{2}\chi_{01}
\left[3\partial_i\partial_j\delta_s-\delta_{ij}\nabla^2\delta_s\right]^2\!\!\!\!(\vx)
\nonumber \;,
\end{align}
provided that we ignore all the contributions involving moments at zero lag. Nonlocality enters
through the filtering solely (which is the reason why we still call it a local expansion).
As emphasized in 
\cite{2010PhRvD..82j3529D}, it is important to realize that $\dpk$ is {\it not} a count-in-cell 
quantity. Counts-in-cells can generally be constructed using the void generating function, see 
e.g. \cite{1975ApJ...196..647P,1979MNRAS.186..145W}, but it is beyond the scope of this paper 
to compute moments of the peak frequency distribution function.
Since all the bias factors are peak-background split biases obtained 
from a suitable average of orthogonal polynomials, one could try to write down an expansion in 
terms of orthogonal polynomials in the variables $\delta_s$, $\grad^2\delta_s$ etc. such that 
all the contributions involving moments at zero lag cancel out. We will explore this possibility
in future work.
Note that this idea was put forward for the first time by \cite{1988ApJ...333...21S}, who 
considered correlations of regions above threshold as a proxy for luminous tracers. More recently, 
\cite{2012arXiv1205.3401M} proposed an algorithm based on Hermite polynomials to extract 
$k$-dependent bias factors from cross-correlations between the halo and Hermite-transformed 
mass density field.

To make connection with the formalism of \cite{2011PhRvD..83h3518M} (see also 
\cite{1995ApJS..101....1M}), we define the Fourier space peak bias parameters 
$c_n(\vk_1,\dots,\vk_n)$ as the sum over all the contributions to $\dpk(\vx)$ from a given 
order. We thus have
\begin{equation}
c_1(\vk) \equiv \left(b_{10} + b_{01} k^2\right) W(kR_s)
\end{equation}
and
\begin{align}
c_2(\vk_1,\vk_2) &\equiv \biggl\{b_{20} + b_{11} \left(k_1^2+k_2^2\right) 
+ b_{02} k_1^2 k_2^2  \\ 
& \qquad -2 \chi_{10} \left(\vk_1\cdot\vk_2\right) +\chi_{01}
\biggl[3\left(\vk_1\cdot\vk_2\right)^2 \nonumber \\ 
& \qquad -k_1^2 k_2^2\biggr]\biggr\}\, W(k_1 R_s) W(k_2 R_s) \nonumber \;,
\end{align}
where $W(kR_s)$ is the smoothing kernel. These definitions are consistent with those of the 
``renormalized'' bias parameters introduced by \cite{2011PhRvD..83h3518M} who argued that, 
owing to rotational symmetry, the peak bias parameters should take the above functional form. 
In particular, the correspondence between the bias factors associated with $\eta^2$ and 
$\zeta^2$ and those of \cite{2011PhRvD..83h3518M} is $E_2=3\chi_{01}$, $C_2=-2\chi_{10}$ and 
$D_2=b_{02}-\chi_{01}$.
We stress, however, that the peak bias factors discussed in this work have not been obtained 
by means of a renormalization procedure. 
We speculate that the local bias expansion Eq.(\ref{eq:newdpk}) can be extended to all orders
to match the exact peak 2-point correlation function at all separations. Namely, 
\begin{align}
\dpk(\vx) &= \sum_{n=1}^\infty \frac{1}{n!}
\int\!\!\frac{d^3k_1}{(2\pi)^3}\dots\frac{d^3k_n}{(2\pi)^n}\,
c_n(\vk_1,\dots,\vk_n) \\
& \qquad \times \delta(\vk_1)\dots\delta(\vk_n) e^{i(\vk_1+\dots+\vk_n)\cdot\vx}
\nonumber \;,
\end{align}
where $c_n(\vk_1,\dots,\vk_n)$ is a sum over all the possible combinations of rotational 
invariants involving exactly $n$ powers of the linear density field $\delta_s$ and/or its 
derivatives.

Our peak-background split approach provides a simple way of predicting the bias coefficients
associated with any rotational invariant quantity. The Hermite weighting scheme introduced 
by \cite{2012arXiv1205.3401M} furnishes a practical way of measuring the biases $b_{ij}$ 
from simulations. Clearly, their scheme could be extended to also measure the biases 
$\chi_{ij}$. However, because discrete density maxima require a somewhat more sophisticated 
treatment of counts-in-cells, we leave this for future work. Here, we merely establish a 
recursion relation between the $b_{ij}$ by considering either the property
\begin{equation}
\left(\frac{\partial}{\partial\nu}+\gamma_1\frac{\partial}{\partial u}\right) e^{-Q_1(\vy)} 
= -\nu e^{-Q_1(\vy)} \;,
\end{equation}
or the generating function in Eq.(\ref{eq:Tnormal}). In the latter case, upon substituting
$b_u=(\sigma_0/\sigma_1)^2(\nu/\sigma_0-b_\nu)$ in the exponential factor, we find 
\begin{equation}
\delta_c b_{01} = \left(\frac{\sigma_0}{\sigma_1}\right)^2 \left(\nu^2-\delta_c b_{10}\right)
\end{equation}
at the first order whereas, at the second order, we obtain
\begin{align}
\delta_c^2 b_{11} &= \delta_c^2\, \ov{b_\nu b_u}+
\frac{\delta_c^2 \gamma_1^2}{(1-\gamma_1^2)} \\ 
&= \left(\frac{\sigma_0}{\sigma_1}\right)^2
\Biggl[\delta_c^2\left(\frac{\nu}{\sigma_0}\right)\ov{b}_\nu-\delta_c^2 \ov{b_\nu^2} 
+\frac{\gamma_1^2\nu^2}{(1-\gamma_1^2)}\Biggr] \nonumber \\
&= \left(\frac{\sigma_0}{\sigma_1}\right)^2 
\Bigl(-\nu^2+\nu^2\delta_c b_{10}-\delta_c^2 b_{20}\Bigr) \nonumber \;,
\end{align}
and
\begin{align}
\delta_c^2 b_{02} &= 
\delta_c^2 \ov{b_u^2} -\frac{\delta_c^2}{\sigma_2^2(1-\gamma_1^2)} \\
&= \left(\frac{\sigma_0}{\sigma_1}\right)^4 
\Biggl[\nu^4-2\delta_c^2\left(\frac{\nu}{\sigma_0}\right)\ov{b}_\nu
+\ov{b_\nu^2} -\frac{\gamma_1^2\nu^2}{(1-\gamma_1^2)}\Biggr] \nonumber \\
&= \left(\frac{\sigma_0}{\sigma_1}\right)^4
\Bigl[\nu^2\left(\nu^2+1\right)-2\nu^2\delta_c b_{10}+\delta_c^2 b_{20}\Bigr] \nonumber \;.
\end{align}
At least for $n=1,2$, $b_{kl}$ with $1\leq l\leq k\leq n$ can be expressed as a linear combination
of $b_{k0}$, $1\leq k\leq n$, plus a polynomial in $\nu$. This agrees with the findings of 
\cite{2012arXiv1205.3401M} obtained within the excursion set approach \cite{2012arXiv1205.3401M}
(except for the multiplicative factors of $(\sigma_0/\sigma_1)^{2l}$). 
Similar relations should hold at any order in the peak bias parameters $b_{ij}$. 
This structure arises from the fact that the peak-background split acts on probability densities, 
which are continuous functions of space. For discrete tracers such as density maxima, the background 
perturbation affects the fields appearing in the Gaussian multivariate $P_1(\vy)$, but not those  
entering the expression of the peak number density $\npk(\vy)$. The peak constraint weights the 
peak-background split series expansion such that the peak bias factors are recovered. 

%%%%%%%%%%%%%%%%%%%%%%%%%%%%%%%%%
\begin{figure*}
\center
\resizebox{0.45\textwidth}{!}{\includegraphics{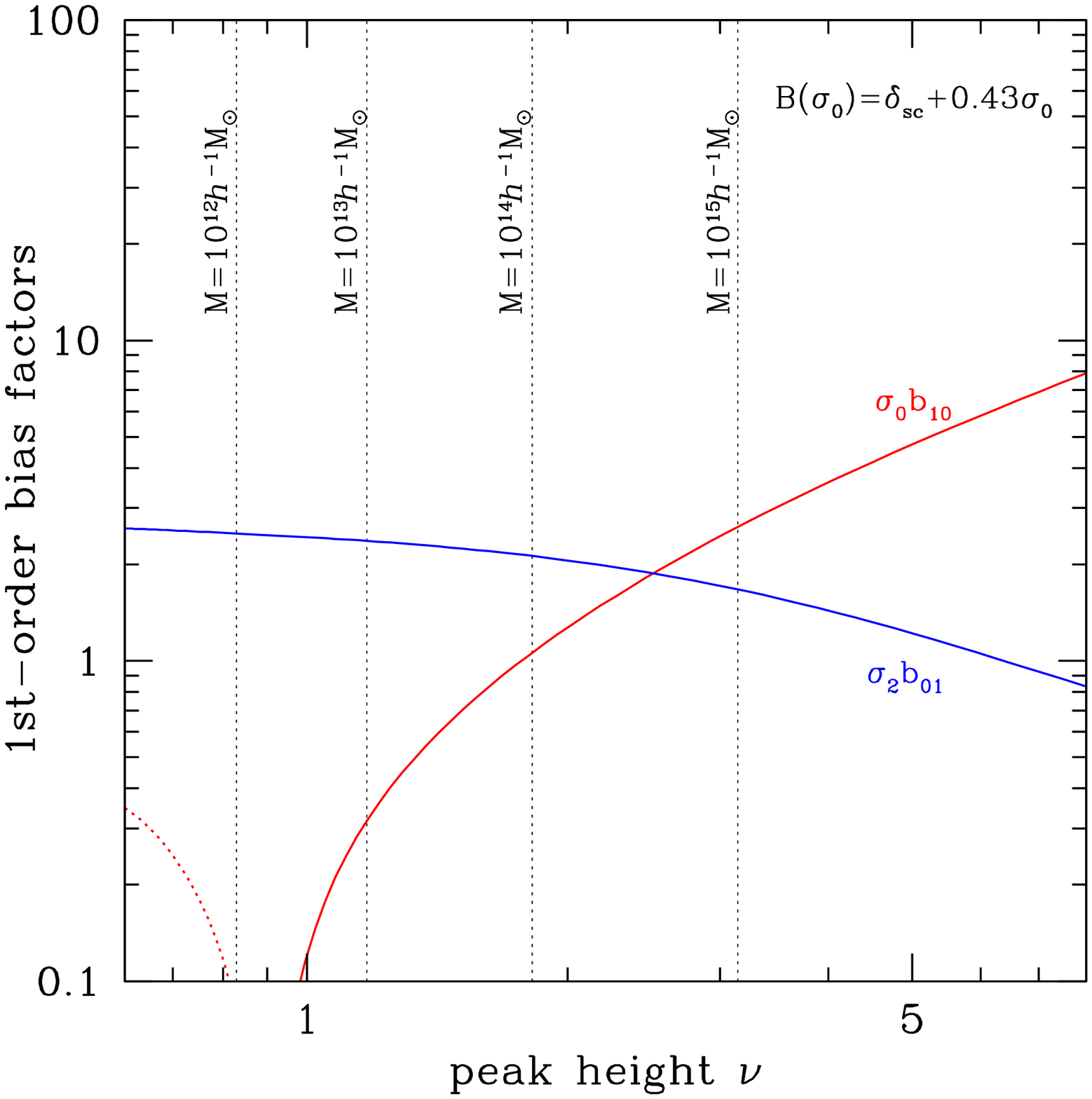}}
\resizebox{0.45\textwidth}{!}{\includegraphics{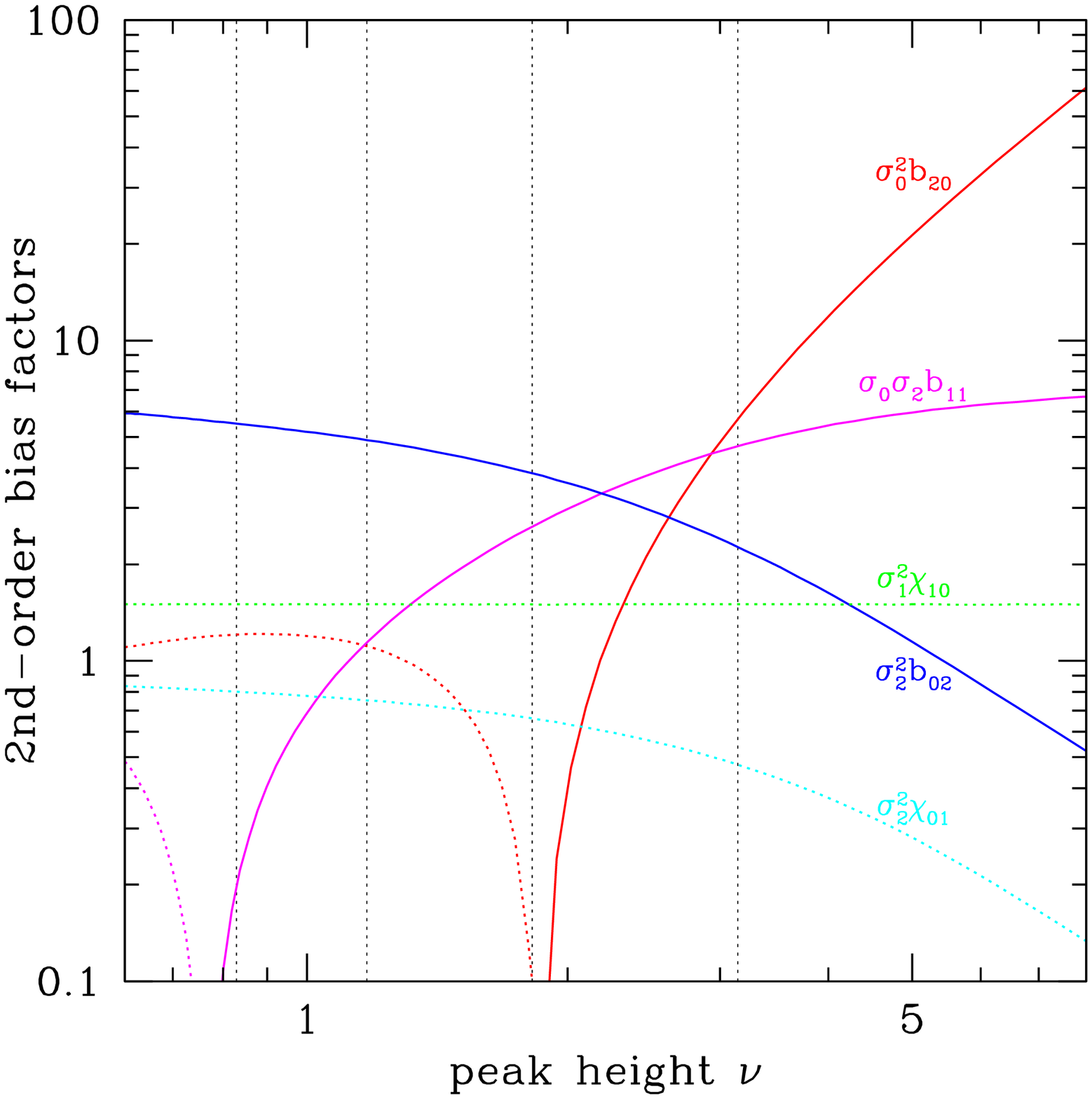}}
\caption{First-order (left panel) and second-order (right panel) Lagrangian peak bias factors
derived from the excursion set peaks mass function Eq.(\ref{eq:espfnu}) assuming a moving barrier 
$B(\sigma_0)=\delta_c+0.43\sigma_0$ (see text). The bias parameters have been multiplied by the
appropriate factors of $\sigma_i$, so that they all are dimensionless and multiply combinations
of normalized variables in the effective local bias expansion Eq.(\ref{eq:newdpk}). 
In particular, the second-order bias factors $\sigma_0^2 b_{20}$, 
$\sigma_0\sigma_2 b_{11}$ and $\sigma_2^2 b_{02}$ are associated with $\nu(\vx)$ and $u(\vx)$,
whereas $\sigma_1^2 \chi_{10}$ and $\sigma_2^2 \chi_{01}$ weight the contributions from  
$\eta^2(\vx)$ and $\zeta^2(\vx)$, respectively. Vertical lines mark the peak significance at which 
the halo mass is $M=10^{12}$, $10^{13}$, $10^{14}$ and $10^{15}\hmsun$ (from left to right). 
Dashed curves indicate negative values. Results are shown at $z=0$ for a $\Lambda$CDM cosmology 
with normalisation $\sigma_8=0.82$.}
\label{fig:bias}
\end{figure*}

\subsection{Correlation functions for excursion set peaks}

To make connection with the clustering of dark matter halos, we must ensure that the density
in a tophat region centered at the peak location never reaches the collapse threshold 
($\delta_c$) on any smoothing scale $R>R_s$.
Ref.~\cite{2012MNRAS.423L.102M} showed that enforcing the conditions $\delta(R_s)>\delta_c$ and 
$\delta(R_s+\Delta R_s)<\delta_c$ as in \cite{1990MNRAS.245..522A} provides a very good 
approximation to the first-crossing distribution when the stochastic walks generated from the 
variation of $R_s$ are strongly correlated. 
An important consequence of this result is the possibility of restricting the excursion set to 
those locations that meet the peak constraint \cite{2012MNRAS.426.2789P}.
The number density of dark matter halos per unit mass and volume is usually written
\begin{equation}
\bar{n}(M) = \frac{\bar{\rho}}{M} f(\nu) \frac{d\nu}{dM}\;,
\end{equation}
where $f(\nu)$ is the multiplicity function. Following the approach of 
\cite{1990MNRAS.245..522A,2012MNRAS.426.2789P}, the number density of peaks identified on the
filtering scale $R_s$ and satisfying the aforementioned conditions is
\begin{multline}
\nesp(R_s) dR_s \\ 
= \frac{3^{3/2}}{R_1^3} \left(\frac{\sigma_2}{\sigma_0}\right)
\int\!\!d^{10}\vy\,\npk(\vy)\, u\, P_1(\vy) R_s dR_s \;.
\end{multline}
where, for simplicity, we have assumed that the smoothing kernel is Gaussian, but it is 
straightforward to generalize these results to arbitrary filters. Therefore,  
$d\nu/dR_s=\nu R_s (\sigma_1/\sigma_0)^2$ and we can write the excursion set peaks multiplicity
function as
\begin{align}
\label{eq:espfnu}
f_{\rm ESP}(\nu) &= \left(\frac{M}{\bar{\rho}}\right)\nesp(R_s)\frac{dR_s}{d\nu} \\
&= \frac{e^{-\nu^2/2}}{\sqrt{2\pi}} \left(\frac{V}{V_\star}\right) 
\frac{G_1^{(1)}\!(\gamma_1,\gamma_1\nu)}{\gamma_1\nu} \nonumber \\ 
&= V \nesp(\nu) \nonumber\;.
\end{align}
Here, $V\equiv M/\bar{\rho}$ is the Lagrangian volume associated with the filter (usually tophat). 
$f_{\rm ESP}$ is the fundamental ingredient in the mass function prediction of 
\cite{2012arXiv1210.1483P} since it can be interpreted as a multiplicity function. 
It is pretty clear that, in the peak 2-point correlation, the constraint $\delta(R_s)>\delta_c$ 
and $\delta(R_s+\Delta R_s)<\delta_c$ will translate into an extra multiplicative factor of 
$(\sigma_2/\sigma_0)u R_s dR_s$ in the integrand. 
Therefore, the ESP correlation functions can also be obtained in the exact same way as that of 
the BBKS peaks, but with a peak number density
\begin{align}
n_{\rm ESP}(\nu',R_s,\vx) &\equiv 
\frac{3^{3/2}}{R_\star^3} \left(-\frac{\tr\zeta}{\gamma_1\nu'}\right)
|\det\zeta|\,
\delta_D\!\left(\veta\right) \\ 
& \quad \times \theta_H\!\left(\lambda_3\right)\delta_D\!\left(\nu-\nu'\right) \nonumber \;.
\end{align}
The bias parameters for excursion set peaks are thus given by
\begin{align}
b_{ij} &= \frac{1}{\sigma_0^i\sigma_2^j\bar{n}_{\rm ESP}}
\int\!\!d^{10}\vy\, n_{\rm ESP}(\vy) H_{ij}(\nu,u)\,P_1(\vy) \\
\chi_{k0} &= \frac{(-1)^k}{\sigma_1^{2k}\bar{n}_{\rm ESP}}
\int\!\!d^{10}\vy\, n_{\rm ESP}(\vy)\,L_k^{(1/2)}\!\!\left(\frac{3\eta^2}{2}\right) P_1(\vy) 
\nonumber \\
\chi_{0k} &= \frac{(-1)^k}{\sigma_2^{2k}\bar{n}_{\rm ESP}}
\int\!\!d^{10}\vy\, n_{\rm ESP}(\vy)\,L_k^{(3/2)}\!\!\left(\frac{5\zeta^2}{2}\right) P_1(\vy) 
\nonumber \;.
\end{align}
They are similar to the bias factors of BBKS peaks, except for the fact that the $k$th order 
moment $\overline{u^k}$ of the peak curvature must be replaced by $\overline{u^{k+1}}/\overline{u}$ 
in Eqs.(\ref{eq:bvv}) -- (\ref{eq:buu}), and $G_0^{(\alpha)}$ must be replaced by $G_1^{(\alpha)}$ 
in Eq.(\ref{eq:bzz}). 
For large values of $\omega=\gamma_1\nu$, $G_1^{(\alpha)}$ asymptotes to 
$G_1^{(\alpha)}\approx \alpha^{-5/2}\omega^4$. This implies that
$\partial_\alpha\ln G_1^{(\alpha)}\!(\gamma_1,\omega)$ converges towards -5/2 in the limit 
$\omega \to \infty$. 
Hence, the second-order bias induced by the asymmetry of the peak profile also vanishes in 
the high peak limit for the ESP multiplicity function, as it should be.

To illustrate the behaviour of the ESP bias parameters, we follow the methodology adopted in 
\cite{2012arXiv1210.1483P}. Namely, we smooth the density field with a tophat filter, while
sticking to the Gaussian filter to define $\eta_i$ and $\zeta_{ij}$ (so that the second 
spectral moment $\sigma_2$ remains finite).
Therefore, we have $\gamma_1=\sigma_{1\times}^2/(\sigma_{0T}\sigma_{2G})$, where the subscript 
$T$ and $G$ refer to tophat and Gaussian filtering, respectively, and $\times$ denotes a 
mixed filtering, i.e. one filter is Gaussian and the other is tophat.
Next, we construct the mapping between $R_T$ and $R_G$ by finding the $R_G(R_T)$ for which 
$\la\delta_G\delta_T\ra=\la\delta_T^2\ra$. 
Finally, to account for departures from the spherical collapse approximation, we consider a 
moving barrier of the (square-root) form $B=\dsc+\beta\sigma_0$, where $\beta=0.43$ 
\citep{2009ApJ...696..636R}. For simplicity, we will ignore the scatter around $B$ even 
though it is quite substantial in the range of $\nu$ we are interested in.
Fig.\ref{fig:bias} shows the first- and second-order peak bias factors at $z=0$ for a 
$\Lambda$CDM cosmology with $\sigma_8=0.82$. Note that, while $M=10^{13}\hmsun$ translate into
a significance of $\nu\approx 1.2$, the actual height of $\nu=1.2$ density peaks is 
$B/\sigma_0\approx 1.6$.

These results can be generalized to arbitrary filtering of the mass density field. For 
non-Gaussian initial conditions, there are a couple of subtleties which will be discussed 
elsewhere~\footnote{V. Desjacques, J.-O. Gong, A. Riotto, in preparation}.

\section{Discussion and Conclusions}
\label{sec:discussion}

We have shown that the 2-point correlation function $\xpk(r)$ of discrete density peaks can 
be computed, up to second order at least, from an {\it effective} local bias expansion in 
continuous fields that are invariant under rotations of the coordinate frame. This local 
expansion is not a count-in-cell relation in the sense that $\dpk(\vx)$ is merely an effective 
overdensity that can be used to recover the true $\xpk(r)$ from a trivial evaluation of 
$\la\dpk(\vx_1)\dpk(\vx_2)\ra$. One of the consequences is that there only is one physically
motivated smothing scale: the Lagrangian radius $R_s$ of the halos. Yet another important 
difference with the widespread local bias model is that one shall ignore all the contributions 
from zero-lag moments in order to obtain the correct $\xpk(r)$. 

All the bias coefficients can be derived from a peak-background split argument in which the 
background perturbation shifts the zero mean of the 1-point probability distribution functions 
of the rotationally invariant fields, unlike essentially all the other peak-background split 
formulations which consider a change in the number density of the tracers. Consequently, it
is possible to derive bias factors from a peak-background split argument even if the variables
are integrated over.
The resulting probability densities can then be expanded in orthogonal polynomial bases. For the 
normally distributed peak height $\nu(\vx)$ and curvature $u(\vx)$, these are bivariate 
Hermite polynomials whereas, for the chi-squared distributed $\eta^2(\vx)$ and $\zeta^2(\vx)$, 
these are generalized Laguerre polynomials. The peak bias factors are then obtained upon 
averaging the appropriate orthogonal polynomials over all the spatial locations that satisfy 
the peak constraint. 

We have demonstrated that our simple local expansion reproduces the 2-point peak correlation 
function $\xpk(r)$ computed at second order by \cite{2010PhRvD..82j3529D} after a tedious 
expansion of the joint probability density $P_2(\vy_1,\vy_2;r)$. 
We believe that it should remain valid at 
higher orders. Furthermore, because discreteness enters the calculation only when averaging the 
orthogonal polynomials, we speculate that this local bias expansion combined with the 
peak-background split approach presented here can be generalized to describe the clustering of 
any point process of a Gaussian random field. The great advantage of our approach is that it 
circumvents the computation of $P_2(\vy_1,\vy_2;r)$, and requires only the evaluation of 
$P_1(\vy)$. 

Our approach can be easily generalized to more sophisticated constraints involving, for instance,
the tidal shear $\partial_i\partial_j\Phi(\vx)$, where $\Phi(\vx)$ is the gravitational potential. 
As noted in \cite{2002MNRAS.329...61S,2012arXiv1207.7117S}, the quadratic invariant $s^2(\vx)$ 
(the equivalent of our $\zeta^2(\vx)$ but with $\delta$ replaced by $\Phi$) follows a 
$\chi^2$-distribution with 5 degrees of freedom. Therefore, we expect that its associated bias
parameters are given by some suitable average of the Laguerre polynomials $L_k^{(3/2)}(x)$. If
the (nonspherical) collapse occurs at the spatial location of density peaks or includes the
dependence on the large scale environment, then the $\chi_5^2$ distribution will be replaced by
the appropriate conditional probability density \cite{2008MNRAS.388..638D,2012MNRAS.421..296R},
to which we shall apply the peak-background split in order to read off the new bias parameters.

Corrections induced by nonlinear gravitational evolution can also be decomposed into rotational
invariants \cite{1992ApJ...394L...5B,1995MNRAS.276...39C,2009JCAP...08..020M}. 
Therefore, if one ignores the diffusion kernels (i.e. the propagators introduced by 
\cite{2006PhRvD..73f3519C}), then it is straightforward to find explicit expressions for the 
Eulerian bias parameters in terms of local and nonlocal Lagrangian bias factors 
\cite{1998MNRAS.297..692C,2012PhRvD..85h3509C,2012PhRvD..86h3540B,2012arXiv1207.7117S}.
This procedure can clearly be applied to our effective bias expansion Eq.(\ref{eq:newdpk}),
with the important caveat that discrete density peaks exhibit a statistical velocity bias.
\cite{2010PhRvD..81b3526D}.
Notwithstanding this, we expect from the structure of the $F_2$ kernel that the bias factors 
$\chi_{ij}$ remain constant with time, in agreement with the findings of 
\cite{2010PhRvD..82j3529D}.
For a more realistic treatment of gravitational motions, it should be possible to compute the 
evolved 2-point peak correlation $\xpk(r,z)$ in the framework of the integrated perturbation 
theory proposed by \cite{2011PhRvD..83h3518M}. We leave all this to future work.

\section*{Acknowledgments}

I would like to thank Matteo Biagetti and Kwan Chuen Chan for their careful reading of the 
manuscript, and the Swiss National Science Foundation for support.

\bibliographystyle{prsty}
\bibliography{references}

\appendix

\section{Computing correlators}
\label{sec:correlators}

For illustation, we evaluate the cross-covariance $\la\eta_1^2\eta_2^2\ra$ of the square 
modulus of the gradient $\eta^2(\vx)$ at two different spatial locations $\vx_1$ and 
$\vx_2$. We have
\begin{align}
\left\la\eta_1^2\eta_2^2\right\ra &= \frac{1}{\sigma_1^4}
\Biggl\{\prod_{i=1}^4\int\!\!\frac{d^3k_i}{(2\pi)^3}\Biggr\}\left(\vk_1\cdot\vk_2\right)
\left(\vk_3\cdot\vk_4\right) \\
& \quad \times
\left\la\delta_s(\vk_1)\delta_s(\vk_2)\delta_s(\vk_3)\delta_s(\vk_4)\right\ra 
e^{i(\vk_1+\vk_2)\cdot\vx_2+i(\vk_3+\vk_4)\cdot\vx_1} \nonumber \\
&= 1 + \frac{2}{\sigma_1^4}\int\!\!\frac{d^3k_1}{(2\pi)^3}\int\!\!\frac{d^3k_2}{(2\pi)^3}\,
\left(\vk_1\cdot\vk_2\right)^2 P_s(k_1) P_s(k_2) \nonumber \\
&\quad \times e^{i(\vk_1+\vk_2)\cdot\vr} \nonumber \\
& = 1 + \left(\frac{2}{\sigma_1^4}\right)\delta_{ij}\delta_{lm}\, 
J_{il}(r) J_{jm}(r) \nonumber \;,
\end{align}
where
\begin{equation}
J_{ij}(r)=\int\!\!\frac{d^3k}{(2\pi)^3} k_i k_j P_s(k) e^{i\vk\cdot\vr} \;.
\end{equation}
To evaluate $J_{ij}(r)$, we express $k_i$ in terms of the components $\kh_i=k_i/k$ of the
unit vector, and take advantage of the fact that the integral over the angular variables is
\begin{multline}
\frac{1}{4\pi}\int\!\!d\Omega_{\kvh}\, \kh_i \kh_j e^{i\vk\cdot\vr} \\
= \frac{1}{3}\Bigl[j_0(kr)+j_2(kr)\Bigr]\delta_{ij}-j_2(kr)\rh_i\rh_j \;.
\end{multline}
Therefore, the product $\delta_{ij}\delta_{lm} J_{il}(r) J_{jm}(r)$ becomes (we omit the 
$r$-dependence for conciseness)
\begin{align}
\delta_{ij}\delta_{lm} J_{ij} & J_{lm} =
\delta_{ij}\delta_{lm}
\biggl[\frac{1}{3}\bigl(\xi_0^{(1)}+\xi_2^{(1)}\bigr)\delta_{il}-\xi_2^{(1)}\rh_i\rh_l\biggr]
\nonumber \\
&\quad \times 
\biggl[\frac{1}{3}\bigl(\xi_0^{(1)}+\xi_2^{(1)}\bigr)\delta_{jm}-\xi_2^{(1)}\rh_j\rh_m\biggr]
\nonumber \\
&= \frac{1}{3}\bigl(\xi_0^{(1)}+\xi_2^{(1)}\bigr)^2
-\frac{2}{3}\xi_2^{(1)}\bigr(\xi_0^{(1)}+\xi_2^{(1)}\bigr)+\bigr(\xi_2^{(1)}\bigr)^2
\nonumber \\
& = \frac{1}{3}\Bigl[\bigl(\xi_0^{(1)}\bigr)^2+2\bigl(\xi_2^{(1)}\bigr)^2\Bigr] \;,
\end{align}
which yields Eq.(\ref{eq:eta1eta2}) once the multiplicative factor of $(2/\sigma_1^4)$ and the 
additive zero-lag contribution are accounted for.

\section{Shape factor for peaks}
\label{sec:uvw}

In terms of the ordered eigenvalues $\lambda_1\geq\lambda_2\geq\lambda_3$ of the hessian 
matrix $\zeta_{ij}$, the asymmetry parameters that quantify the departure from a spherically
symmetric peak density profile are $v=(\lambda_1-\lambda_3)/2$ and 
$w=(\lambda_1-2\lambda_2+\lambda_3)/2$. 
The peak constraint together, with our choice of ordering, impose the four conditions 
$v\geq 0$, $-v\leq w\leq v$, $(u+w)\geq 3v$ and $u\geq 0$. 
Following \cite{1986ApJ...304...15B}, we also introduce an auxiliary function that 
measures the degree of asphericity expected for a peak, 
\begin{eqnarray}
F(u,v,w) = 
\left(u-2w\right)\Bigl[\left(u+w\right)^2-9v^2\Bigr]v\left(v^2-w^2\right)
\end{eqnarray}
This function scales as $\propto u^3$ in the limit $u\gg 1$. 

In \cite{2010PhRvD..82j3529D}, the peak 2-point correlation up to second order (i.e. 
terms quadratic in the correlation of the density field and its derivatives) is written
as the sum of the linear contribution and three second-order terms $\xi_{\rm pk}^{(2i)}$,
$i=1,2,3$. In particular, $\xpk^{(22)}$ contains all the terms for which the 
$\nu$-dependence cannot be expressed as a polynomial in the linear and quadratic bias 
parameters $b_{ij}$. Their expression is phrased in terms of 
\begin{widetext}
\begin{multline}
\label{eq:fua}
f(u,\alpha) \equiv \frac{3^2 5^{5/2}}{\sqrt{2\pi}}
\left\{\int_0^{u/4}\!\!d v \int_{-v}^{+v}\!\!d w
+\int_{u/4}^{u/2}\!\!d v \int_{3v-w}^v\!\!d w\right\} F(u,v,w)\, 
e^{-\frac{5\alpha}{2}\left(3 v^2+w^2\right)} \\
= \frac{1}{\alpha^4}\Biggl\{\frac{e^{-5\alpha u^2/2}}{\sqrt{10\pi}}
\left(-\frac{16}{5}+\alpha u^2\right)+\frac{e^{-5\alpha u^2/8}}
{\sqrt{10\pi}}\left(\frac{16}{5}+\frac{31}{2}\alpha u^2\right) \\
+\frac{\sqrt{\alpha}}{2}\left(\alpha u^3-3u\right)
\left[{\rm Erf}\left(\sqrt{\frac{5\alpha}{2}}\frac{u}{2}\right)+
{\rm Erf}\left(\sqrt{\frac{5\alpha}{2}}u\right)\right]\Biggr\} \;,
\end{multline}
\end{widetext}
and its integral over the $n$th power of the peak curvature $u$ times the $u$-dependent 
part of the one-point probability distribution,
\begin{equation}
G_n^{(\alpha)}(\gamma_1,w)=\int_0^\infty\!\!d x\, x^n f(x,\alpha) 
\frac{e^{-(x-w)^2/2(1-\gamma_1^2)}}{\sqrt{2\pi\left(1-\gamma_1^2\right)}}\;.
\label{eq:Gn}
\end{equation}
These functions are very similar, albeit more general than those defined in Eqs (A15) 
and (A19) of \cite{1986ApJ...304...15B}.

\section{Non-central chi-squared distributions}
\label{sec:chisquare}

The probability density of a non-central chi-squared distribution $\chi_k^{'2}(x;\lambda)$ 
with $k$ degrees of freedom and non-centrality parameter $\lambda$ is given by
\begin{equation}
\chi_k^{'2}(x;\lambda) = \frac{e^{-(x+\lambda)/2}}{2} 
\left(\frac{x}{\lambda}\right)^{\alpha/2} I_\alpha\left(\sqrt{\lambda x}\right)\;,
\end{equation}
where $\alpha=k/2-1$ and $I_\alpha(x)$ is a modified Bessel function of the 
first kind. 
Ref.~\cite{tiku:1965} proposed the following Laguerre polynomial expansion,
\begin{align}
\chi_k^{'2}(x;\lambda) &= 
\frac{e^{-x/2}}{2} \left(\frac{x}{2}\right)^\alpha
\sum_{j=0}^\infty \frac{\left(-\frac{\lambda}{2}\right)^j}
{\Gamma\left(\frac{1}{2}k+j\right)}\,
L_j^{(\alpha)}\!\!\left(\frac{x}{2}\right) \\
&= \chi_k^2(x) \sum_{j=0}^\infty 
\frac{\Gamma\bigl(\frac{1}{2}\nu\bigr)}{\Gamma\bigl(\frac{1}{2}\nu+j\bigr)}
\left(-\frac{\lambda}{2}\right)^j L_j^{(\alpha)}\!\!\left(\frac{x}{2}\right)
\nonumber \;.
\end{align}
The non-central $\chi^2$-distribution can also be represented as a Poisson-weighted 
mixture of central $\chi^2$-distributions (this was used by, e.g., \cite{2012arXiv1207.7117S}
to estimate the nonlocal Lagrangian bias induced by ellipsoidal collapse). Note, however, 
that this representation does not make apparent the connection with the bias parameters 
$\chi_{ij}$.

\end{document}